\documentclass[a4paper,11pt]{article}
\usepackage{jheppub} 
\usepackage{float}

\title{\boldmath The boundary entropy function for interface conformal field theories}

\author[a,b]{Evangelos Afxonidis,}
\author[b]{Andreas Karch,}
\author[c]{and Chitraang Murdia}

\affiliation[a]{Department of Physics and Insituto de Ciencias y Technolog\'ias Especiales de Asturias (ICTEA), Universidad de Oviedo, c/ Leopoldo Calvo Sotelo 18, ES-33007, Oviedo, Spain}
\affiliation[b]{Weinberg Institute, Department of Physics, University of Texas, 2615 Speedway, Austin, TX 78712, USA}
\affiliation[c]{Department of Physics and Astronomy, University of Pennsylvania, Philadelphia, PA 19104, USA}

\emailAdd{afxonidisevangelos@uniovi.es}
\emailAdd{karcha@utexas.edu}
\emailAdd{murdia@sas.upenn.edu}

\abstract{In 1+1 dimensional conformal field theory with a boundary the boundary contribution to the entanglement entropy is determined by a single number $g$ effectively counting the boundary degrees of freedom. In contrast, in 1+1 dimensional interface CFTs the corresponding quantity is a non-trivial {\it function} depending on the position of the interval relative to the interface, giving access to much more detailed information about the defect. In this work we determined this $g$-function in several examples using holography and derive some of its basic properties from holography and strong subadditivity.}

\begin{document}
\maketitle
\flushbottom

\section{Introduction}
\label{sec:intro}
The foundational work of Cardy catalyzed the study of Boundary Conformal Field Theories (BCFTs) \cite{Cardy:1986gw}, establishing profound connections with both condensed matter physics and string theory. More recently, Interface Conformal Field Theories (ICFTs) have garnered attention as well \cite{Bachas:2001vj}, albeit to a lesser extent. An ICFT describes a system in which a conformally invariant interface joins two, potentially distinct, Conformal Field Theories (CFTs) with differing central charges.

ICFTs can be viewed as a specialized extension of BCFTs, preserving analogous symmetries. Utilizing the folding trick, acting with $x \to -x$ on the left half-space, one can transform the ICFT setup into a BCFT framework.
While this is formally an identification, certain physical quantities that were natural before folding appear atypical in the folded picture and hence, have escaped attention when focusing on the BCFTs.

One simple to understand example of such a quantity is the energy transmission coefficient $T$ \cite{Quella:2006de, Meineri:2019ycm}, which clearly is defined only in the case of interfaces, and not for boundaries. In this work, we focus on another such ICFT-specific quantity, the entanglement entropy (EE) associated with an interval $\mathcal{A}$ crossing the interface. This interval has a total length $l$, with the portions of the interval on either side of the interface with lengths $l_L$ and $l_R$, respectively.

Such an interface crossing interval $\mathcal{A}$ can be folded into an interval terminating on the BCFT boundary
provided $l_L$ and $l_R$ are equal.  Within this context, the entanglement entropy (EE) can be computed using the replica method \cite{Calabrese:2009qy}, yielding
\begin{equation}
\label{bcftee}
    S_\mathcal{A}=\frac{c_L+c_R}{6}\log{\frac{l}{\epsilon}}+\log{g} \ ,
\end{equation}
where $l=l_L+l_R$ denotes the total interval length, $c_L$ and $c_R$ are the central charges on either side of the interface, so that $c_L + c_R$ is the central charge of the BCFT after folding, and $\epsilon$ represents the UV cutoff.

The quantity $\log{g}$ appearing in \eqref{bcftee} is a pivotal quantity characterizing defects or boundaries and is referred to as the boundary entropy \cite{Calabrese:2009qy}, which had previously been introduced using a thermodynamic definition in \cite{Affleck:1991tk}. 
This quantity measures the degrees of freedom localized at a defect and serves as a boundary analogue of the central charge $c$. While $g$ is known in many free or solvable examples, calculating the boundary entropy in strongly coupled systems is challenging. However, holography provides a powerful tool for this purpose by enabling the computation of the boundary entropy via the holographic entanglement entropy (HEE) prescription \cite{Ryu:2006ef}. 

In the case of a generic asymmetric ICFT interval, $g$ in \eqref{bcftee} gets replaced with an effective
$g_{\text{eff}}$-function \cite{Azeyanagi:2007qj, Karch:2000gx}, which depends non-trivially on the ratio $l_L/L_R$. Only for $l_L = l_R$ does $g_{\text{eff}}$ reduce to the standard $g$, $g_{\text{eff}}(1) = g$. The degenerate case with either $l_L$ or $l_R$ vanishing is even more intriguing, in that case even the coefficient of the log divergent term changes, and one of the central charges in \eqref{bcftee} gets replaced by an effective central charge $c_{\text{eff}}$ \cite{Sakai:2008tt, Brehm:2015lja, Karch:2000gx}. This effective central charge has played a crucial role in much recent work on this subject \cite{Karch:2022vot, Karch:2023evr, Karch:2024udk}, where it was shown that it can be thought of as a transmission coefficient for quantum information that obeys interesting bounds, especially when compared to the transmission coefficient for energy. In contrast, in this work we will stay away from this extreme limit and instead focus on $g_{\text{eff}}$, which governs the behavior of the EE at finite $l_L$ and $l_R$.

In the case of $l_L \neq l_R$, the same EE can be expressed in two distinct forms as follows
\begin{align}{\label{eq::EE1_diffc}}
    S_{\mathcal{A}} &= \frac{c_L+c_R}{6}\log{\frac{l_L+l_R}{\epsilon}}+\log g_{\text{eff}}^{(1)} \ , \\
\label{eq::EE2_diffc}
    S_{\mathcal{A}} &= \frac{c_L}{6}\log \frac{2l_L}{\epsilon}+\frac{c_R}{6}\log \frac{2l_R}{\epsilon}+\log g_{\text{eff}}^{(2)} \ .
\end{align}
In the first case, the leading piece corresponds to the standard EE of an ICFT, whereas in the second case, the leading piece corresponds to the sum of EEs for two decoupled BCFTs.
Note that these two equations represent the same physical quantity in two different schemes, and we can convert between them using the relationship 
\begin{equation}
\label{eq:geff1geff2conv}
    \log g_{\text{eff}}^{(1)}= \log g_{\text{eff}}^{(2)} + \frac{c_L}{6} \log \frac{2 l_L/l_R}{\left(1 + l_L/l_R \right)} + \frac{c_R}{6} \log \frac{2}{\left(1 + l_L/l_R \right)}\ .
\end{equation}
In either scheme the precise form of $\log g_{\text{eff}}$ depends on the details of the ICFT.
It is worth reiterating that for $l_L = l_R$, the scheme dependence goes away and $ g_{\text{eff}}^{(1)} = g_{\text{eff}}^{(2)} = g$.

In this work, we will focus entirely on the somewhat special case where the central charge $c$ on the two sides of the interface is the same, $c_L = c_R = c$. Very little is known about $g_{\text{eff}}$ even in this limited context. It has previously been analyzed in a few explicit holographic examples such as the Janus and RS braneworld solutions \cite{Karch:2021qhd, Chapman:2018bqj, Anous:2022wqh}, or in free fermion theories \cite{Kruthoff:2021vgv}. These studies revealed that $g_{\text{eff}}$ indeed depends intricately on the ratio $l_L/l_R$ of the intervals around the interface. We will assume that $g_{\text{eff}}$ depends on the lengths $l_L$ and $l_R$ via only this ratio $l_L/l_R$ for the rest of this paper. In this case, the two different expressions for the EE are
\begin{align}
    {\label{eq::EE1}}
    S_{\mathcal{A}} &= \frac{c}{3}\log{\frac{l_L+l_R}{\epsilon}}+\log g_{\text{eff}}^{(1)}(l_L/l_R) \ , \\
\label{eq::EE2}
    S_{\mathcal{A}} &= \frac{c}{6} \log \frac{4 l_L l_R}{\epsilon^2} + \log g_{\text{eff}}^{(2)}(l_L/l_R) \ 
\end{align}

The 3d Janus solution \cite{Bak:2007jm}, a straightforward generalization of the original 5d Janus solution constructed in \cite{Bak:2003jk}, is a fully backreacted type IIB supergravity solution asymptotically approaching $AdS_3\times S^3 \times T^4$.\footnote{The compactification manifold $M_4$ may alternatively be $T^4$ or $K_3$, but $T^4$ offers a simpler dual SCFT.} 
It is holographically dual to an interface theory where the conformal field theory on either side of the interface is a non-linear sigma model with a $T^4/S_N$ target space. 
The marginal coupling corresponding to the $T^4$ size modulus remains constant in the two $1+1$-dimensional half-spaces but exhibits a discontinuity across a planar $0+1$-dimensional space. The subcritical Randall-Sundrum (RS) braneworld model \cite{Karch:2000gx, Karch:2000ct}, by contrast, is based on Einstein gravity with a negative cosmological constant and includes a brane as matter content. 
For brane tensions below a critical value, the brane intersects the boundary along a timelike defect, with the regions on either side of the defect being empty AdS spaces. This setup provides a holographic dual to ICFTs, albeit as a bottom-up construction without explicit derivation from string theory.

In this holographic setting, a distinctive feature of $g_{\text{eff}}$ is its sensitivity to the full geometry of the solution as encoded in the warp factor $e^{A(r)}$, as extensively analyzed in the context of the super-Janus solution in this paper. Super-Janus solutions are once again type IIB supergravity configurations asymptotic to $AdS_3\times S^3\times T^4$, and, like 3d Janus itself, are dual to marginal deformations of two-dimensional $\mathcal{N}=(4,4)$ SCFTs. Unlike the simpler Janus solution, the super-Janus metric involves a fibration over a Riemann surface, complicating its dimensional reduction to three dimensions \cite{Baig:2024hfc}. From the point of view of the CFT, extra terms are added on the interface to ensure that it preserves half of the supersymmetries. In contrast, in the standard 3d Janus solution, the interface, while preserving conformal invariance, breaks all supersymmetries.

To conduct our analysis, we employ HEE tools \cite{Ryu:2006bv, Rangamani:2016dms, Ryu:2006ef}. The AdS/CFT correspondence, first proposed in \cite{Maldacena:1997re}, posits an equivalence between gravity in a $d+1$-dimensional AdS spacetime and a $d$-dimensional CFT \cite{Maldacena:1997re, Witten:1998qj}. Within this framework, the EE $S_{\mathcal{A}}$ of an entangled region $\mathcal{A}$ is geometrically computed via the area of an extremal surface $\Gamma_\mathcal{A}$, denoted as $|\Gamma_\mathcal{A}|$, homologous to $\mathcal{A}$ and terminating on its boundary in AdS:
\begin{equation}\label{eq::RangamaniEE}
    S_\mathcal{A}=\frac{|\Gamma_\mathcal{A}|}{4G} \ ,
\end{equation}
where $G$ is the Newton constant.
We provide holographic computations for several cases involving interfaces whose bulk theories exhibit conformal invariance.

While HEE tools provide us with interesting examples where we can compute the interface entropy, we use results from quantum information theory to demonstrate general properties. A cornerstone of quantum information theory is the strong subadditivity inequality (SSA) \cite{Lieb:1973zz}, expressed as
\begin{equation}\label{eq::SSA}
    S_A + S_B \geq  S_{A\cap B} + S_{A\cup B} \ .
\end{equation}
HEE offers a geometrical derivation of SSA, rooted in the triangle inequality of Euclidean geometry \cite{Headrick:2007km}. Furthermore, SSA underpins the derivation of $g$-theorems in BCFTs and ICFTs \cite{Harper:2024aku}. 
In this paper, we use SSA to derive a $g_{\text{eff}}$-theorem for the interface entropy $g_{\text{eff}}^{(2)}$. We also obtain another theory-dependent constraint on $g_{\text{eff}}^{(2)}$ as well as equivalent constraints on $g_{\text{eff}}^{(1)}$. 
We verify that these bounds are satisfied by the holographic and free fermion theories.

This paper is organized as follows. Section \ref{sec::examples} studies $g_{\text{eff}}$ in three examples -- the super-Janus, Janus, and RS braneworld solutions -- emphasizing its dependence on the spacetime warp factor. 
In section \ref{sec::gtheorem}, we derive the constraints on $g_{\text{eff}}^{(1),(2)}$ using SSA and verify that these constraints are satisfied by various examples.
In appendix \ref{Appendix::KK}, we present technical details, in particular the dimensional reduction of the boundary entropy $g$ and the effective central charge $c_{\text{eff}}$ under KK compactification.

\section{Holographic examples of \texorpdfstring{$g_{\text{eff}}(l_L/l_R)$}{interface entropy}}
\label{sec::examples}

\subsection{Setup}
Holographic ICFTs are often described by three-dimensional gravity duals. Consistency within string theory suggests that these duals typically correspond to ten-dimensional spacetimes with seven compactified dimensions. In some cases, the internal space is non-trivially fibered over the relevant three-dimensional spacetime, as seen in the gravitational duals of $3+1$ dimensional supersymmetric Janus solutions introduced in \cite{DHoker:2007zhm}. The starting point for the construction of conformal defect spacetimes in $2+1$ dimensions is given by a metric of the form
\begin{equation}\label{eq::AdS2slicingsofAdS3}
    ds^2=e^{A(r)}\frac{dx^2-dt^2}{x^2}+dr^2 \ .
\end{equation}
This configuration can be interpreted as $AdS_2$ slices within an $AdS_3$ spacetime. The two-dimensional slices depend on the holographic direction through the warp factor $A(r)$. The case of pure $AdS_3$ corresponds to the presence of a completely transparent defect on the $1+1$ dimensional boundary and its warp factor is given by 
\begin{equation}
    e^{A(r)}=\cosh{(r)} \ .
\end{equation}
The boundary of $AdS_3$ consists of two copies of $AdS_2$ which are approached as $r\rightarrow\pm\infty$, with an interface bridging the two regions. A convenient description that will occasionally be employed is provided by the Poincaré coordinates, in which the boundary is located at $z=0$
\begin{equation}\label{eq::Poincarepatch}
    ds^2=\frac{1}{z^2}\left( -dt^2+dy^2+dz^2\right) \ .
\end{equation}
The two coordinate systems (\ref{eq::AdS2slicingsofAdS3}) and (\ref{eq::Poincarepatch}) are related by the following transformations
\begin{equation}\label{eq::PoincarePatchCoordChange}
    z=\frac{x}{\cosh{r}},\quad\quad y=x\tanh{r} \ .
\end{equation}

In this paper, we focus on the case where the intervals on the left and right of the interface are unequal. This disparity can be encoded in the integration constant $c_s$ arising from the scale isometry, $x\rightarrow \lambda x$, of the $AdS_2$ slices.

Consider $l_L\equiv \lim_{r\rightarrow -\infty} x(r)$ the interval lying to the left of the interface, and $l_R\equiv \lim_{r\rightarrow \infty} x(r)$ the interval on the right. To compute the minimal RT surface, as prescribed in \cite{Ryu:2006ef}, we define the Lagrangian as follows
\begin{equation}
    \mathcal{L}=\sqrt{\frac{e^{2A}}{x^2}(x')^2+1} \ .
\end{equation}
The associated Noether charge is
\begin{equation}
    c_s=\frac{\partial \mathcal{L}}{\partial x'}\frac{\partial(\lambda x)}{\partial \lambda}=\frac{e^{2A} x'}{\sqrt{x^2+e^{2A}(x')^2}} \ .
\end{equation}
which solving for $x'$ yields
\begin{equation}\label{eq::scalesymmetryLagrSolution}
    \frac{x'}{x}=\pm \frac{c_s e^{-A}}{\sqrt{e^{2A}-c_s^2}} \ .
\end{equation}
We can immediately obtain the constraint 
\begin{equation}
    0\leq |c_s|\leq e^{A_*} \ ,
\end{equation}
where $e^{A_*}$ is the minimal warp factor. Since (\ref{eq::scalesymmetryLagrSolution}) has a sign ambiguity nothing is lost by restricting to positive $c_s$. For the case of pure $AdS_3$ case we have $e^{A_*}=1$. Without loss of generality we can choose the positive sign in (\ref{eq::scalesymmetryLagrSolution}). The on-shell Lagrangian for the positive solution of (\ref{eq::scalesymmetryLagrSolution}) is given by
\begin{equation}\label{eq::onshellLagrangian}
    \mathcal{L}=\frac{1}{\sqrt{1-c_s^2e^{-2A}}} \ .
\end{equation}

Depending on what value $c_s$ takes, we can distinguish different cases for the extremal RT surface:
\begin{itemize}
\item For $c_s=0$, we get from (\ref{eq::scalesymmetryLagrSolution}) that $x'=0$ and hence $x=l_R=\text{const}$, due to the asymptotics at $r\rightarrow \infty$. That is the case where $l_R=l_L=l/2$, i.e. the interval around the interface is symmetric. This case corresponds to the standard BCFT result, where the standard folding trick yields $S=\frac{2c}{6}\log{(l/\epsilon)}$ for the entanglement entropy.
\item For $0<c_s<e^{A_*}$, we have that $1/\sqrt{1-c_s^2e^{-2A}}$ is always positive and thus we fall into the case of $l_R\geq l_L$ if we chose the plus sign in (\ref{eq::scalesymmetryLagrSolution}), since $x'$ is strictly positive in this case, and $l_L \geq l_R$ if we chose the minus sign. Either way this realizes the case of a non-trivial ICFT setup. Notice that the folding trick cannot be usefully applied here. The non-symmetric interval would fold to a configuration with 3 regions, where all degrees of freedom are traced out in one, some in a second, and none in the third. This is not a standard setup for  EE. Without loss of generality, we limit ourselves to the case that $l_R \geq l_L$.

\item When $c_s$ takes the maximum value $c_s=e^{A_*}$ the solution truncates at $r=0$ and we get not just two but actually {\it four} different solutions depending on the choice of sign in (\ref{eq::scalesymmetryLagrSolution}) and whether we look at positive or negative $r$. Focusing on positive $r$ and choosing the plus sign in  (\ref{eq::scalesymmetryLagrSolution}) yields the case of $l_L=0$. This is because, (\ref{eq::scalesymmetryLagrSolution}) diverges at $r\rightarrow 0$ unless $x=0$ at that point. The plus sign in (\ref{eq::scalesymmetryLagrSolution}) allows for a solution which terminates at $x(0)=0$ with $x$ smoothly increasing to $x(\infty)=l_R$. This solution corresponds to an RT surface ending on the interface with $l_L=0$. The solution with the minus sign will play a prominent role when we revisit the case of the bosonic Janus.
\item For $c_s > e^{A_*}$ one finds RT surfaces which have two endpoints on the same side of the interface and so are of no interest for our study.
\end{itemize}
Therefore, we see that the value of $c_s$ affects the length of the intervals around the interface and thus the EE computation.

\subsection{Super-Janus and non-universality of \texorpdfstring{$g_{\text{eff}}$}{interface entropy} }\label{sec::superJanus}
In this section we study the holographic dual of the 2D supersymmetric Janus ICFT, originally constructed in \cite{Chiodaroli:2009yw}. It arises as a solution of type IIB supergravity, which is locally asymptotic to $AdS_3 \times S^3 \times M_4$. The compactification manifold $M_4$ can be either $T^4$ or $K_3$, corresponding to the target space of the dual CFT. For simplicity we consider the case of $T^4$.

The ten-dimensional metric is a fibration of $AdS_2\times S^2\times T^4$ over a Riemann surface $\Sigma$ and given by
\begin{equation}\label{eq::superJanusmetric}
    ds^2_{(10)}=f_1^2 ds^2_{AdS_2}+f_2^2 ds^2_{S^2}+f_3^2 ds^2_{T^4}+\rho^2 ds^2_{\Sigma} \ ,
\end{equation}
which contains five parameters $\theta,\psi, L, k$ and $b$. $\psi$ parametrizes the jump of the six-dimensional dilaton through the interface, while $\theta$ the jump of the axion. The various functions in the metric above are given by two meromoprhic function and two harmonic functions \cite{Chiodaroli:2010ur}. Note that setting $\theta=0$ and $\psi=0$ gives the $AdS_3 \times S^3$ vacuum.

The goal is to reduce the ten-dimensional system down to a three-dimensional geometry. One can achieve this by first compactifying (\ref{eq::superJanusmetric}) over $T^4$ down to six dimensions. The resulting metric is of the form $AdS_2\times S^2 \times \Sigma$ and be further reduced to three dimensions by first compactifying on $S^2$ and then on the compact part of $\Sigma$. Notice that all functions depend on the complex coordinate on the fibration $\Sigma$ and one needs to be attentive when KK reducing \cite{Baig:2024hfc}. The resulting three-dimensional metric is given by
\begin{equation}\label{eq::KKsuperJanusmetric}
    ds^2_{(3)}=R^2\left( \frac{\cosh^2{(r+\psi)}}{\cosh^2{\psi}\cosh^2{\theta}} ds^2_{AdS_2} +dr^2\right)=R^2\left( e^{2A(r)}ds^2_{AdS_2}+dr^2\right) \ .
\end{equation}
with the $AdS_3$ radius and the warp factor given by
\begin{equation}\label{eq::AdS3radiuspluswarpfactor}
    R^2=2L\cosh{\psi}\cosh{\theta},\quad \quad e^{A(r)}=\frac{\cosh({r+\psi)}}{\cosh{\psi}\cosh{\theta}} \ .
\end{equation}

Following \cite{Karch:2021qhd} we can find that the integration constant $c_s$ takes the following values
\begin{equation}
    0\leq c_s \leq R \, e^{A_*} = \sqrt{\frac{2L}{\cosh{\psi}\cosh{\theta}}} \ .
\end{equation}
Furthermore, for the case of asymmetric intervals around the interface, $0<c_s<Re^{A_*}$, substituting the super Janus warp factor (\ref{eq::AdS3radiuspluswarpfactor}) into the equations of motion (\ref{eq::scalesymmetryLagrSolution}) yields
\begin{equation}\label{eq::superJanuseom}
    \frac{x(r)}{x_0}=\exp{\left[\pm \tanh^{-1}{\left( \frac{c_s \sinh{(r+\psi)}}{\sqrt{-c_s^2+\frac{2L\cosh^2{(r+\psi)}}{\cosh{\psi}\cosh{\theta}}}}\right)}\cosh{\theta}\cosh{\psi} \right]} \ .
\end{equation}
By considering the  $x\rightarrow \pm \infty$  limits of (\ref{eq::superJanuseom}), one can determine
\begin{align}
\label{eq::superJanuslR}
   l_R&=x_0\exp{\left[ \tanh^{-1}{\left( \frac{c_s}{\sqrt{\frac{2L}{\cosh{\theta}\cosh{\psi}}}} \right)}\cosh{\theta}\cosh{\psi}\right]} \ , \\
\label{eq::superJanuslL}
    l_L&=x_0\exp{\left[ -\tanh^{-1}{\left( \frac{c_s}{\sqrt{\frac{2L}{\cosh{\theta}\cosh{\psi}}}} \right)}\cosh{\theta}\cosh{\psi}\right]} \ .
\end{align}

To compute the regulated area $\mathcal{A}$, we must integrate the on-shell Lagrangian, incorporating the super-Janus warp factor. Analogous to \cite{Karch:2021qhd}, the asymptotic value of the super-Janus warp factor (\ref{eq::AdS3radiuspluswarpfactor}) takes the form
\begin{equation}\label{eq::asymptoticsuperJanus}
    e^A\sim\frac{1}{2\cosh{\theta}\cosh{\psi}}e^r=\frac{x}{z} \ .
\end{equation}
Therefore, the regulated limits are given by
\begin{equation}\label{eq::regulatedlimitsSuperJanus}
    r_c^{\pm}=\log{\left(\frac{2l_{R/L}}{\epsilon}\right)}+\log{(\cosh{\psi}\cosh{\theta})} \ .
\end{equation}
The regulated limits are consistent, as the $g$-number for the super Janus configuration can be computed as 
\begin{align}
    \mathcal{A}_{c_s=0}=\int_{-x_c^-}^{x_c^+}R(\mathcal{L}=1)=2R\log{\left( \frac{l}{\epsilon}\right)}+2R\log{\cosh{\psi}\cosh{\theta}} \ .
\end{align}
This aligns with the 3D calculation of the $g$-number performed in appendix \ref{Appendix::KK} and the 6D analysis presented in \cite{Chiodaroli:2010ur}. 

\begin{figure}[H]
\centering
\includegraphics[width=0.9\textwidth]{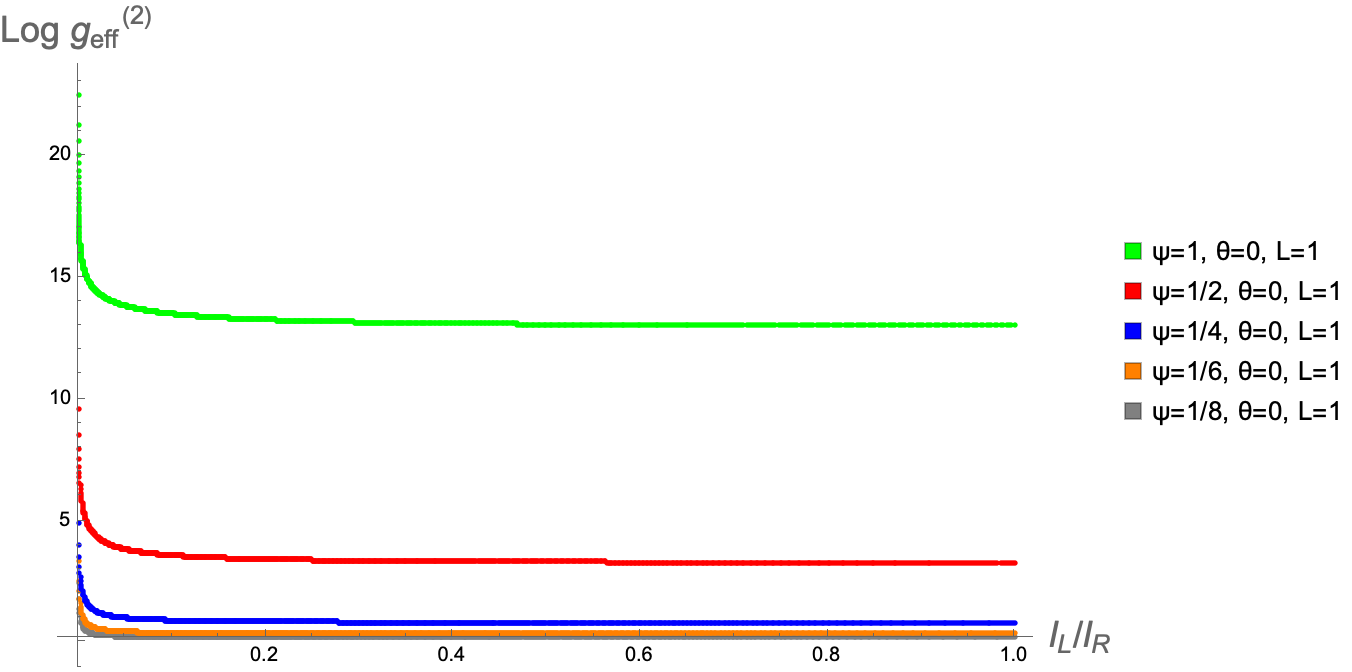}
\caption{Effective interface entropy $\log g_{\text{eff}}^{(2)}$ for the super-Janus solution as a function of the ratio $l_L/l_R$. \label{fig:LoggeffvsratioSuperJanusPaper}}
\end{figure}

For $0<c_s<\sqrt{\frac{2L}{\cosh{\psi}\cosh{\theta}}}$, the regulated area $\mathcal{A}$ is determined through numerical integration of the on-shell Lagrangian (\ref{eq::onshellLagrangian}) with the super-Janus warp factor. For arbitrary values of  $\psi$ and $\theta=0$ in the super-Janus solution with $L=1$, the result is shown in figure \ref{fig:LoggeffvsratioSuperJanusPaper}.

The interface entropy $\log{g_{\text{eff}}}^{(2)}$ is computed as the difference between the minimal RT surface in the super-Janus case and the empty AdS, $\mathcal{A}(\theta=0,\psi\neq 0)-\mathcal{A}(0,0)$, similar to \cite{Azeyanagi:2007qj}. Crucially, all curves diverge when the ratio approaches zero, indicating that the interval becomes maximally asymmetric. This is in contrast to the Janus solution as in \cite{Karch:2021qhd}, where the curves blow up for different values of the ratio. We will analyze the cause of this difference in the next section. On the other hand, the curves attain their boundary entropy values as the ratio approaches one.

\begin{figure}[H]
\centering
\includegraphics[width=1\textwidth]{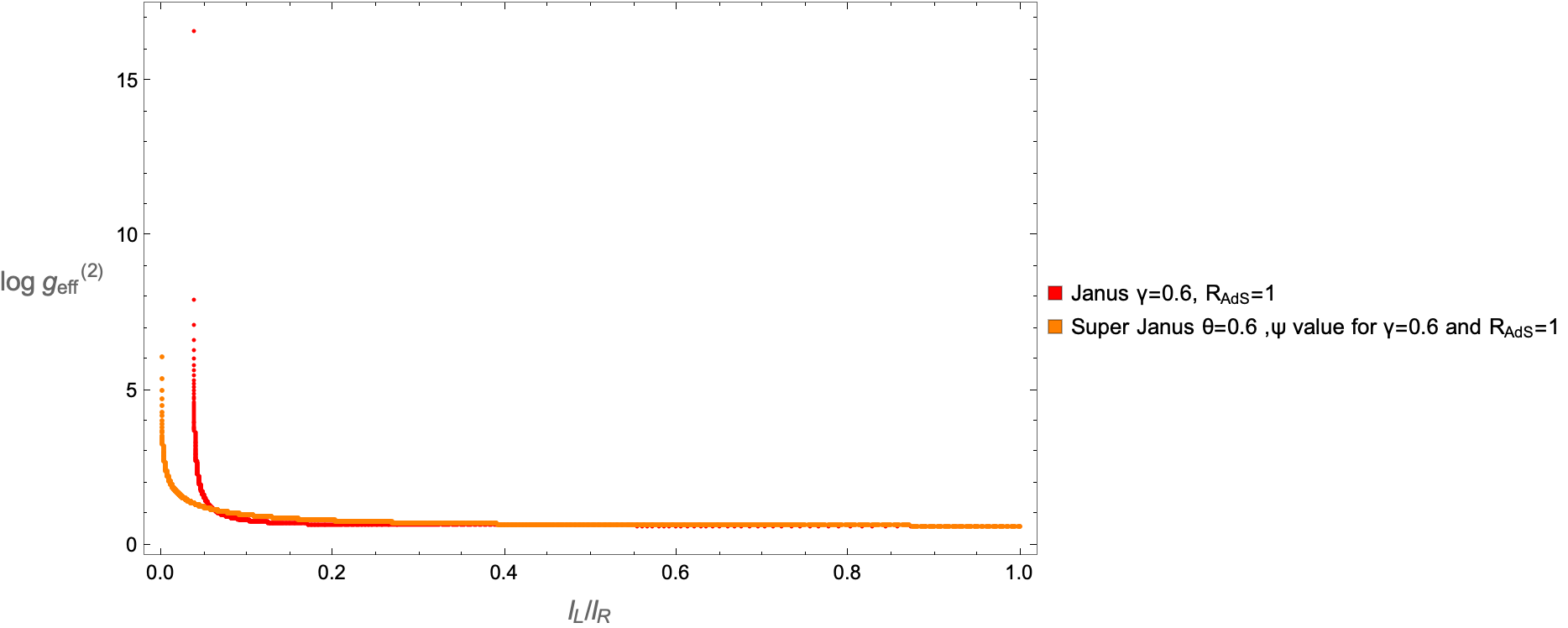}
\caption{Effective interface entropy as a function of the ratio $l_L/l_R$ for both the Janus and super-Janus solution. Note that both curves have the same boundary entropy $g$ but they diverge for different values of the ratio. \label{fig:loggeffvsratioUNIVERSALITYPAPER}}
\end{figure}

An intriguing check would be to determine whether the interface entropy is universal, in that one may contemplate that the function $g_{\text{eff}}(l_L/l_R)$ is fixed by conformal invariance in terms of the symmetric value at $l_L=l_R$. To investigate this, we can compare the interface entropies of the Janus \cite{Karch:2021qhd} and super-Janus solutions as in figure \ref{fig:LoggeffvsratioSuperJanusPaper}, and assess whether the corresponding curves coincide for the same boundary entropy. Recall that the boundary entropy of the Janus solution is given by \cite{Azeyanagi:2007qj}
\begin{equation}
    \log{g}=\frac{1}{\sqrt{1-2\gamma^2}} \ .
\end{equation}
Therefore, the relation between the two boundary entropies is given by
\begin{equation}
    \cosh^2{\psi}=\frac{1}{\sqrt{1-2\gamma^2}} \ ,
\end{equation}
where we have set $\theta=0$ and $R=1$. We would like to compare the interface entropy $\log{g_{\text{eff}}}$ for the two solutions as a function of the field theory quantity $l_L/l_R$ on the boundary. It is noteworthy that while $\log{g_{\text{eff}}}^{(2)}$ can be plotted analytically in terms of the integration constant $c_s$, mapping $c_s$ to the ratio $l_L/l_R$ is highly non-trivial for both the Janus and super-Janus solution. Consequently, we employ numerical methods to perform this mapping, as illustrated in figure \ref{fig:loggeffvsratioUNIVERSALITYPAPER}.
It is evident that the curves diverge as the interval approaches maximal asymmetry. Throughout this analysis, we have assumed $l_R\geq l_L$, such that the ratio spans the range from 0 to 1. Notably, both curves converge to the same boundary entropy $g$ on the right-hand side, in agreement with established results \cite{Chiodaroli:2010ur}. This consistency renders them a suitable basis for examining the non-universality of the effective entropy at the interface $\log{g_{\text{eff}}}$ for the Janus and super-Janus solutions. We thus conclude that the interface entropy is non-universal, as it is explicitly sensitive to the full warp factor $e^{A(r)}$. In other words, while the value of $g$ at $l_L=l_R$ is fixed, the functional dependence of $g_{\text{eff}}(l_L/l_R)$ is not governed by conformal invariance but rather by the details of the underlying theory.

\subsection{Updates on standard Janus}
\label{sec::upadtesonJanus}

In this section, we provide updates on the interface entropy of the Janus solution previously discussed in \cite{Karch:2021qhd}. The Janus solution was originally introduced as a 3+1 dimensional ICFT with its asymptotically $AdS_5 \times S^5$ supergravity dual in \cite{Bak:2003jk}, with the 1+1 dimensional version dual to an interface in asymptotically $AdS_3 \times S^3 \times T^4$ presented in \cite{Bak:2007jm}. This non-supersymmetric Janus solution is notably simpler than its supersymmetric counterpart, as the three-dimensional metric appears as a direct product factor in the ten-dimensional metric. The compactified three-dimensional part is described by the Einstein-Hilbert action plus a scalar field $\phi$ and given by
\begin{equation}\label{eq::Janusmetric}
    ds_{(3)}^2=dr^2+\frac{1}{2}\left( 1+\sqrt{1-2\gamma^2}\cosh{(2r)}\right)ds^2_{AdS_2} \ ,
\end{equation}
where $|\gamma|\leq 1/\sqrt{2}$. The dilaton is given by \cite{Chiodaroli:2010ur}
\begin{equation}
    \phi(r)=\phi_0+\frac{1}{\sqrt{2}}\log{\left( \frac{1+\sqrt{1-2\gamma^2}+\sqrt{2}\gamma \tanh{r}}{1+\sqrt{1-2\gamma^2}-\sqrt{2}\gamma \tanh{r}}\right)} \ .
\end{equation}
When $\gamma=0$, the dilaton becomes constant $\phi_0$ and the metric reduces to pure $AdS_3$. When $\gamma=1/\sqrt{2}$, the spacetime is simply given by $\mathbb{R}\times AdS_2$. Let us define $\xi\equiv\sqrt{1-2\gamma^2}$. Then, the Janus warp factor is 
\begin{equation}\label{eq::JanusWarpfactor}
    e^{A(r)} = \sqrt{ \frac{1+\xi\cosh{2r}}{2}} \ .
\end{equation}
The integration constant $c_s$ is then bounded by
\begin{equation}
    0\leq c_s\leq \sqrt{\frac{1+\xi}{2}} \ .
\end{equation}

\begin{figure}
\centering
\includegraphics[width=0.9\textwidth]{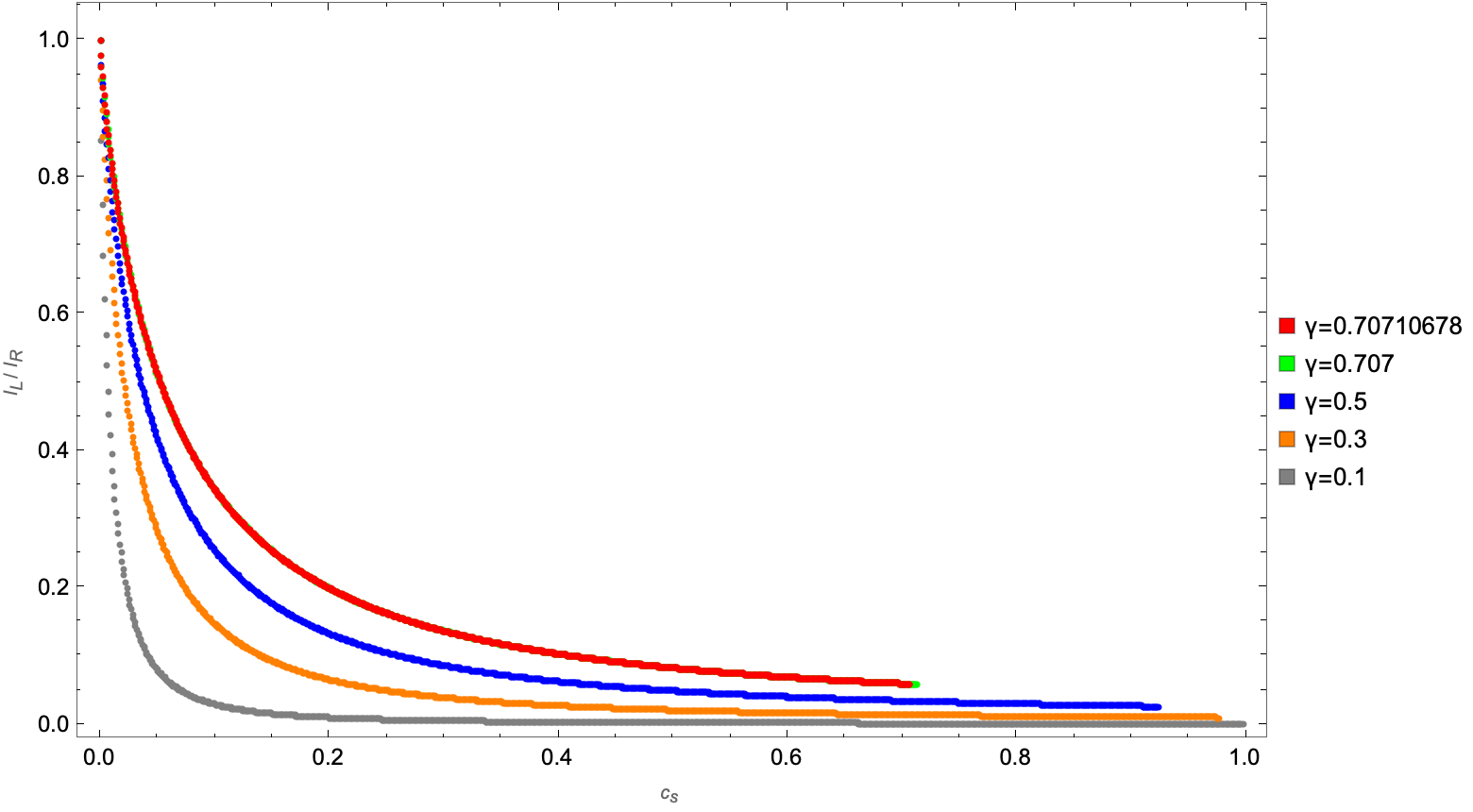}
\caption{Plot of $l_L/l_R$ with respect to the integration constant $c_s$. Notice that the curves acquire a gap in that the terminate at some value of $c_s$ for every given $\gamma$.  \label{fig::ratiovscsPAPER}}
\end{figure}

As discussed in section \ref{sec::superJanus}, a significant distinction arises between the interface entropy profiles of the Janus and super-Janus cases. Specifically, the curves corresponding to different values of $\gamma$ as analyzed in \cite{Karch:2021qhd}, exhibit divergence at distinct values of the ratio $l_L/l_R$. This behavior can be interpreted from a disconnected solution. Plotting the ratio $l_L/l_R$ against the integration constant $c_s$ one obtains figure \ref{fig::ratiovscsPAPER}.

The curves develop a gap, for example, the curve for $\gamma=0.70710678$ terminates at a finite value of the ratio rather than approaching zero. This behavior occurs precisely at the same value of the ratio, where the curves in \cite{Karch:2021qhd} exhibit a divergence. This observation suggests the emergence of disconnected RT surfaces that dominate for smaller values of $l_L/l_R$ and larger values of $c_s$.

The relevant disconnected solution is also given by $c_s=e^{A_*}$, just like the $l_L=0$ case, but this time with the minus sign in (\ref{eq::scalesymmetryLagrSolution}). In this case $x(0)$ can not vanish like it did in the case of an RT surface ending on the interface, since for this sign choice it has to be {\it larger} than $x(\infty)$. The divergence in (\ref{eq::scalesymmetryLagrSolution}) instead will force $x$ to diverge as $r \rightarrow 0$. To verify this, one can substitute this specific value of $c_s$ into the differential equation (\ref{eq::scalesymmetryLagrSolution}). The solution describes an RT surface that starts at the boundary at $r\rightarrow \infty$, $x=l_R$ and then reaches straight down into the bulk towards the Poincare patch horizon. To describe the full RT surface we need a second such horizon crossing surface attached to $x=l_L$ at $r=-\infty$: the full RT surface is disconnected.

We can explicitly write down these disconnected solutions. Solving for $x$ yields
\begin{equation}\label{eq::soldisconnected}
    x(r)=x_0 \sqrt{\frac{1+\chi}{\chi-1}}e^{\frac{1}{\sqrt{\xi}}},\quad \chi\equiv\frac{\sqrt{\xi +1} \cosh (r)}{\sqrt{\xi  \cosh (2 r)+1}} \ .
\end{equation}
which indeed has the properties that 
\begin{align}
    \lim_{r\rightarrow 0^+}x(r)&\longrightarrow \infty , \nonumber\\
    \lim_{r\rightarrow \infty}x(r) &= e^{\frac{i \pi }{2 \sqrt{\xi }}} \left(\frac{\sqrt{2} \sqrt{\frac{1}{\xi }+1}-2}{\sqrt{2} \sqrt{\frac{1}{\xi }+1}+2}\right)^{\frac{1}{2 \sqrt{\xi }}} =l_R \ .
\end{align}
This disconnected solution starts to kick in after $l_L/l_R$ reaches the gap as seen in figure \ref{fig::ratiovscsPAPER}, and subsequently dominates the connected solution given by $c_s=-e^{A}$. 

This disconnected solution leads to a logarithmic IR divergence near the Poincare Patch horizon. To understand this behavior, we examine the on-shell Janus Lagrangian near $r=0$ where $x \rightarrow \infty$. From \eqref{eq::onshellLagrangian} we see that the integrand of the area functional indeed diverges near $r=0$ when $c_s e^{-2 A_*} =1$. To find the behavior near the singularity we can expand the warp factor near $r=0$:
\begin{equation}
A(r) = A_* + \frac{1}{2} A''(0) r^2 + {\cal O} (r^3)  \ .
\end{equation}
and so the integrand of the on-shell action goes as
\begin{equation}
    {\cal L} \sim \sqrt{\frac{2}{A''(0)}} \frac{1}{r} \ .
\end{equation}
Clearly this gives a logarithmically divergent contribution to the area of the disconnected RT surface. One should note that this is really a large $x$ and hence IR divergence. The total length of the geodesic reaching towards the Poincare patch horizon picks up an infinite length from the region hear the horizon in addition to the standard UV divergence from near the boundary. As such, the disconnected solution always has a larger area than the connected solution when the latter exists, as the connected solution is IR finite. In the Standard Janus solution these disconnected solutions have to fill the gap in which the connected solution does not exist. In this regime the EE is very large and sensitive to the IR cutoff.

\subsection{Asymptotic behavior of solutions}
In this section, we investigate the behavior of the Janus and super-Janus solutions as the ratio $l_L/l_R$ approaches its critical blow-up value or zero.

\subsubsection{Janus}
We can expand the solution near $c_s=0$ to compute the next-to-leading order correction in boundary entropy $g$. To do this, we substitute the Janus warp factor into the on-shell Lagrangian and expand for small $c_s$. By integrating and applying the regularized limits, we obtain
\begin{equation}
\begin{split}
    \mathcal{A}&= \left. \left[r+\frac{c_s^2 \tan ^{-1}\left(\frac{(\xi -1) \tanh (r)}{\sqrt{\xi ^2-1}}\right)}{\sqrt{\xi ^2-1}}+\mathcal{O}\left(c_s^3\right)\right]\right\vert_{-r_c^-}^{r_c^+} \\
    &=2 \log \left(l/\epsilon\right)-\log \xi +\frac{c_s^2}{\sqrt{1-\xi^2}}\log{\left[\frac{(\xi-1)\left(\frac{2l^2}{l^2+\epsilon^2 \xi}-1\right)-\sqrt{1-\xi^2}}{-(\xi-1)\left(\frac{2l^2}{l^2+\epsilon^2 \xi}-1\right)-\sqrt{1-\xi^2}}\right]}+\mathcal{O}\left(c_s^3\right) \ .
\end{split}
\end{equation}
The first two terms are in agreement with \cite{Karch:2021qhd}, while the last serves as a correction for small $c_s$. 
It can be easily checked that the last term is positive and real, so the interface entropy is still positive and real.

\subsubsection{Super-Janus}
Let us now turn to the super-Janus solution. In this case, we have found that for $0<c_s<e^{A_*}$, there are connected solutions that yield minimal RT surfaces. Furthermore, one can easily verify that no disconnected solutions exist when $c_s=\pm e^{A_*}$, in contrast to the Janus solution as discussed in section \ref{sec::upadtesonJanus}. It is worth noting that the regulated limits $r_c^{\pm}$ given by (\ref{eq::regulatedlimitsSuperJanus}), have the same structure as those used in pure AdS setups, with the only difference being the additional $\psi$ and $\theta$ dependent terms. This suggests that the super-Janus solution is a shifted version of pure AdS. Consequently, this immediately implies that the variation of the minimal area with respect to $l_L$ and $l_R$ follows a similar pattern
\begin{equation}
    \delta \mathcal{A}=\mathcal{L}\vert_{r=x_c^+}\frac{\delta x_c^+}{\delta l_R}\delta l_R-\mathcal{L}\vert_{r=-x_c^-}\frac{\delta (-x_c^-)}{\delta l_L}\delta l_L \ ,
\end{equation}
which yields
\begin{equation}
    \frac{\delta x_c^{\pm}}{\delta l_{R/L}}=\frac{1}{l_{R/L}} \ ,
\end{equation}
resulting in
\begin{equation}
    \mathcal{A}=\log{\left( \frac{l_L}{\epsilon}\right)} + \log{\left( \frac{l_R}{\epsilon}\right)} + \text{const}=2\log{\left(\frac{l}{\epsilon} \right)} + \text{const}' \ .
\end{equation}
This minimal area gives rise to the universal logarithmic term of the EE. The terms denoted ``constant" are $\epsilon$-independent, but they do depend on the ratio of $l_L/l_R$.

We also aim to examine the asymptotic behavior of the super-Janus solution near $c_s=0$ and at the critical value where the curves exhibit a divergence. Starting with the former, we can expand the solution around $c_s=0$ to compute the next-to-leading order correction to the boundary entropy for the super-Janus solution. To do so, we substitute the super-Janus warp factor into the Lagrangian and perform an expansion for small $c_s$. After integrating and applying the regularized limits, the minimal area is found to be
\begin{equation}
\begin{split}
    \mathcal{A} &=\left(r+\frac{1}{2} c_s^2 \cosh ^2\theta  \cosh ^2\psi  \tanh (r+\psi )+\mathcal{O}\left(c_s^3\right)\right)\bigg\vert_{-r_c^-}^{r_c^+} \\
    &=2\log \left(\frac{l}{\epsilon }\right)+2\log (\cosh \theta \cosh \psi ) \\
    & \hspace{1in} + c_s^2 \cosh^2{\theta}\cosh^2{\psi}\left(\frac{ l^2}{\epsilon ^2 e^{2 \psi }+l^2} -\frac{ \epsilon ^2}{\epsilon ^2+l^2 e^{2 \psi }}\right)+\mathcal{O}\left(c_s^3\right) \ .
\end{split}
\end{equation}

Similar to the Janus case, the first two terms are in agreement with \cite{Chiodaroli:2010ur}, while the latter serves as a correction for small $c_s$. It can, once again, be easily verified that the latter term is both positive and real, ensuring that the interface entropy remains positive and real.

Finally, we perform an expansion near the blow-up point. In this vicinity, we express $c_s$ as $e^{A_*}-\delta$, where $\delta$ is small.  Expanding the minimal area for small $\delta$, we find a divergence of the form
\begin{equation}
    \mathcal{A} \sim f\left(\log{\delta}\right) \ ,
\end{equation}
which indicates the divergence as $\delta \rightarrow 0$, signaling the blow-up behavior, as seen in figure \ref{fig:LoggeffvsratioSuperJanusPaper}.

\subsection{The RS braneworld}
\label{sec::RS}

One example extensively studied in \cite{Karch:2021qhd} is the subcritical Randall-Sundrum (RS) braneworld \cite{Karch:2000gx, Karch:2000ct}. This model, based on Einstein gravity with a negative cosmological constant and matter in the form of a brane, is a simple solution in general relativity. The brane is a thin relativistic sheet with a constant energy density, characterized by its tension. For tensions below a critical value, the brane intersects the boundary along a time-like defect, with both sides of the defect being empty AdS. This setup allows the brane to be interpreted as a holographic dual of an ICFT, though as a bottom-up construction, the dual CFT cannot be explicitly derived from string theory. 

For our purposes, we can view the RS braneworld as a particular instance of the warp factor
\begin{equation}\label{eq::RSwarpfactor}
    e^{A(r)}=\cosh{(r-r_*)} \ .
\end{equation}
The solution is defined piecewise for positive and negative values of $r$, where $r=0$ is the location of the brane and $r_*$ is the turnaround point, which is is uniquely determined by the brane's tension $T$ \cite{Takayanagi:2011zk} as
\begin{equation}
    \log{g}=\frac{2r_*}{4 G}=\frac{c}{3}\tanh^{-1}{\left(\frac{T}{2}\right)} \ .
\end{equation}

\begin{figure}
\centering
\includegraphics[width=0.9\textwidth]{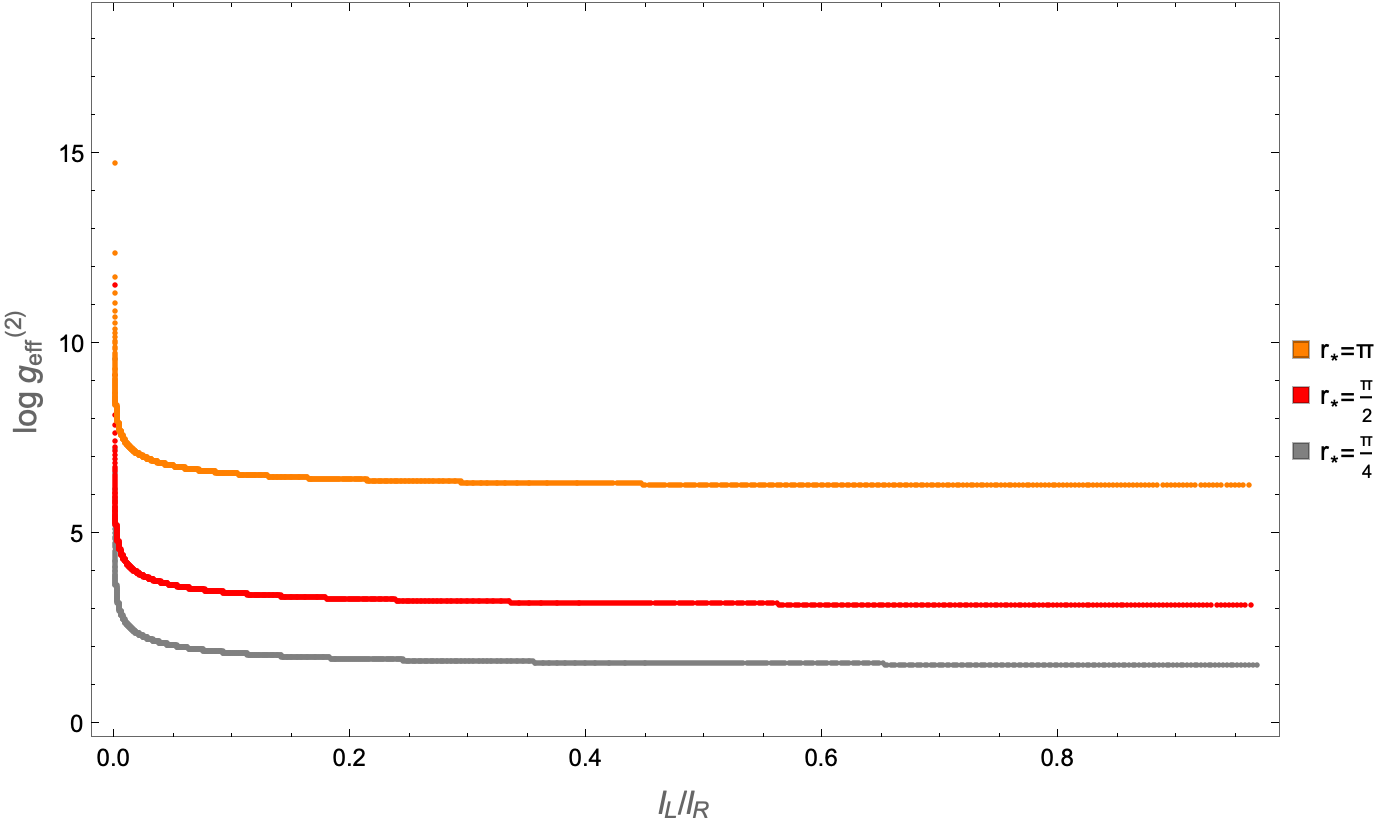}
\caption{Effective interface entropy with respect to the ratio $l_L/l_R$, for different values of the turnaround point $r_*$. Once again, $\log{g_{\text{eff}}}^{(2)}$ is monotonically decreasing as a function of the ratio.  \label{fig::loggeffvsratioRSPAPER}}
\end{figure}

The interface entropy was computed in \cite{Karch:2021qhd} and given by
\begin{equation}
    \log{g_{\text{eff}}}^{(2)}(c_s)=\frac{c}{3}\tanh^{-1}{\frac{\sinh{r_*}}{\sqrt{\cosh^2{r_*}-c_s^2}}} \ .
\end{equation}
Note that the connection between $c_s$ and the ratio $l_L/l_R$ is highly non-trivial. Therefore, one can numerically integrate the on-shell action for the RS warp factor (\ref{eq::RSwarpfactor}). The subleading interface entropy is plotted in figure \ref{fig::loggeffvsratioRSPAPER} with respect to the ratio $l_L/l_R$.

The curves blow up when the ratio becomes zero, meaning that the interval becomes maximally asymmetric and they acquire the $g$-number values when $l_L=l_R$. It is important to note that all the previously discussed examples—the Janus, super-Janus, and RS braneworld solutions—result in a monotonically decreasing interface entropy $\log g_{\text{eff}}^{(2)}$. We provide more details on this in section \ref{sec::gtheorem}.

\subsection{ICFT holographic entanglement and interface entropy}

We return to the general conformal defect metric of the form (\ref{eq::AdS2slicingsofAdS3}).
In this case, the on-shell Lagrangian yields the minimal area 
\begin{equation}
\label{eq::onshellarea}
    \mathcal{A}=\int \frac{e^{A(r)}}{\sqrt{e^{2A(r)}-c_s^2}}dr \ .
\end{equation}
Due to the divergence of the pure AdS warp factor $e^{A(r)}=\cosh{r}$ on the asymptotically AdS  boundaries $r=\pm \infty$ we have to regulate the integration limits as in \cite{Karch:2021qhd}
\begin{equation}
\label{eq::rcpm_logs}
    r_c^+=\log{\left(\frac{2l_R}{\epsilon}\right)},\quad\quad r_c^-=\log{\left( \frac{2l_L}{\epsilon}\right)}\ .
\end{equation}
Then, the on-shell area (\ref{eq::onshellarea}) is given by
\begin{align}
    \mathcal{A}&=\int_{-r_c^-}^{r_c^+} dr\frac{e^{A(r)}}{\sqrt{e^{2A(r)}-c_s^2}}\nonumber \\
    &=\int_{-r_c^-}^\eta dr\frac{e^A(r)}{\sqrt{e^{2A(r)}-c_s^2}}+\int_{\eta}^{r_c^+} dr\frac{e^{A(r)}}{\sqrt{e^{2A(r)}-c_s^2}} \nonumber \\
    &\approx\int_{-r_c^-}^{\eta} dr\frac{e^{-r}/2}{\sqrt{e^{-2r}/4-c_s^2}}+\int_{\eta}^{r_c^+} dr\frac{e^{r}/2}{\sqrt{e^{2r}/4-c_s^2}} \ .
\end{align}
By expanding the on-shell area for large negative and positive values of $r$, it is straightforward to identify that the divergent terms exhibit a logarithmic dependence, and are given by
\begin{equation}
    \mathcal{A}_{\text{div}}=\log{\left(\frac{2l_L}{\epsilon}\right)}+\log{\left(\frac{2l_R}{\epsilon}\right)} \ .
\end{equation}
In this setting, the interface entropy $g_\text{eff}^{(2)}$ which is given in (\ref{eq::EE2}) is more natural because the divergent terms are naturally subtracted off resulting in the well-defined expression
\begin{equation}
\label{eq::interfaceentropy}
    \log{g_\text{eff}^{(2)}}(c_s)
    = \lim_{\epsilon\rightarrow0}\left[ \frac{1}{4 G_N}\left(\int_{-r_c^-}^{r_c^+}\frac{e^{A(r)}}{\sqrt{e^{2A(r)}-c_s^2}}dr-\log{\left(\frac{2l_L}{\epsilon}\right)}-\log{\left(\frac{2l_R}{\epsilon}\right)}\right)\right] \ .
\end{equation}
Notice that the mapping of $c_s$ to the ratio $l_L/l_R$ is highly non-trivial and depends on the specific details of the theory. In the symmetric case, $c_s=0$, the result in (\ref{eq::interfaceentropy}) simplifies to the boundary entropy expression derived in \cite{Gutperle:2012hy}.

Moreover, we can find a general form relating the integration constant $c_s$ and the ratio $l_L/l_R$. To establish this, we begin by considering the differential equation given in (\ref{eq::scalesymmetryLagrSolution}), which can be reformulated as
\begin{equation}
    \frac{d \log{x(r)}}{dr}=\frac{c_s e^{-A(r)}}{\sqrt{e^{2A(r)}-c_s^2}} \ .
\end{equation}
Integrating both sides yields
\begin{equation}\label{eq::relationcsandratio}
    \frac{l_L}{l_R}=\exp{\left(-\int_{-\infty}^{\infty}\frac{c_s e^{-A(r)}}{\sqrt{e^{2A(r)}-c_s^2}}dr\right)} \ .
\end{equation}
Unlike the integral for the area, this integral converges and so we do not need to regulate it and can take the integration limits all the way to infinity.
This result not only shows the dependence of $c_s$ on the geometry but also provides an intuition on how the ratio $l_L/l_R$ changes with respect to $c_s$.

One notable case where we can write down an analytical relation between the integration constant $c_s$ and the ratio $l_L/l_R$ is the super Janus solution. As demonstrated, for $0<c_s<e^{A_*}$, the intervals on either side of the interface are given by (\ref{eq::superJanuslR}) and (\ref{eq::superJanuslL}). Solving for $c_s$, we obtain the following relation
\begin{equation}\label{eq::relationcsratioSJanus}
    c_s=\frac{ \tanh \left[\frac{1}{2}  \log \left(l_R/l_L\right)\text{sech}(\theta ) \text{sech}(\psi )\right]}{\cosh{\theta}\cosh{\psi}} \ ,
\end{equation}
for general values of $\theta$ and $\psi$.

\section{Constraints on \texorpdfstring{$g_{\text{eff}}(l_L/l_R)$}{interface entropy}}
\label{sec::gtheorem}

It is particularly compelling to explore whether fundamental properties of entanglement entropy can yield valuable insights into the behavior of interface entropy and the effective central charge. One question one can explore is whether there is a constraint on the evolution of the effective quantities under RG flow when we deform the ICFT by a relevant deformation. In this case the length of the interval serves as a stand in for RG scale. This strategy was first employed in \cite{Casini:2004bw} to derive the standard $c$-theorem of 2d CFTs. 
More generally, the interface entropy bounds should reduce to boundary entropy constraints as discussed in \cite{Azeyanagi:2007qj}. We use a similar strategy to obtain constraints on the functional form of $g_{\text{eff}}(l_L/l_R)$ for the ICFT.
We propose a $g_{\text{eff}}$-theorem for the interface entropy as well as bounds on $g_{\text{eff}}^{(1)}$. 
These are further examined and verified within the framework of holography and free fermion theory.

The derivation employs the strong subadditivity (SSA) inequality \cite{Casini:2004bw, Hirata:2006jx}, one of the most significant constraints satisfied by entanglement entropy, given by
\begin{equation}
    S_A+S_B \geq S_{A\cap B} + S_{A\cup B} \ .
\end{equation}
For a holographic derivation of SSA see \cite{Hirata:2006jx, Headrick:2007km}.

As explained in section \ref{sec:intro}, the precise definition of the interface entropy is scheme dependent.
There are two distinct natural definitions we employ, $\log g_{\text{eff}}^{(1)}$ and $\log g_{\text{eff}}^{(2)}$ given in (\ref{eq::EE1}) and (\ref{eq::EE2}) respectively.

Let us consider a situation where we couple two identical BCFTs living on the left half space $(-\infty,0]$ and right half-space $[0,\infty)$ via a semi-transparent interface.
For a fully transmissive interface $\log g_{\text{eff}}^{(1)}$ vanishes, whereas for a fully reflective interface it becomes negative. In contrast, $\log g_{\text{eff}}^{(2)}$ is positive for a fully transmissive interface. 

We can understand this behavior qualitatively by perturbing away from the extreme limits. Starting from an ICFT with a fully transmissive interface, the interface entropy is zero because there is no degree of freedom (DoF) localized on the interface. As the interface becomes more reflective, it reduces the entanglement between the left and right sides of the interface. Consequently, the entanglement entropy $S_{\mathcal{A}}$ decreases due to a decrease in the entanglement between the regions $(-\infty, -l_L]$ and $[0,l_R]$ as well as between the regions $[-l_L,0]$ and $[l_R, \infty)$. On the other hand, when we have two decoupled BCFTs with a fully reflective interface, there is no entanglement between the left and right sides. As the interface becomes more transmissive, the entanglement entropy $S_{\mathcal{A}}$ increases due to an increase in the entanglement between the regions $(-\infty, -l_L]$ and $[0,l_R]$ as well as between the regions $[-l_L,0]$ and $[l_R, \infty)$.

\subsection{Entropic \texorpdfstring{$g_{\text{eff}}\,$}{g-effective}-theorem and bounds on \texorpdfstring{$g_{\text{eff}}^{(2)}$}{g-effective-2}}
\label{sec::geff2theorem}

\begin{figure}
    \centering
    \includegraphics[width=0.7\linewidth]{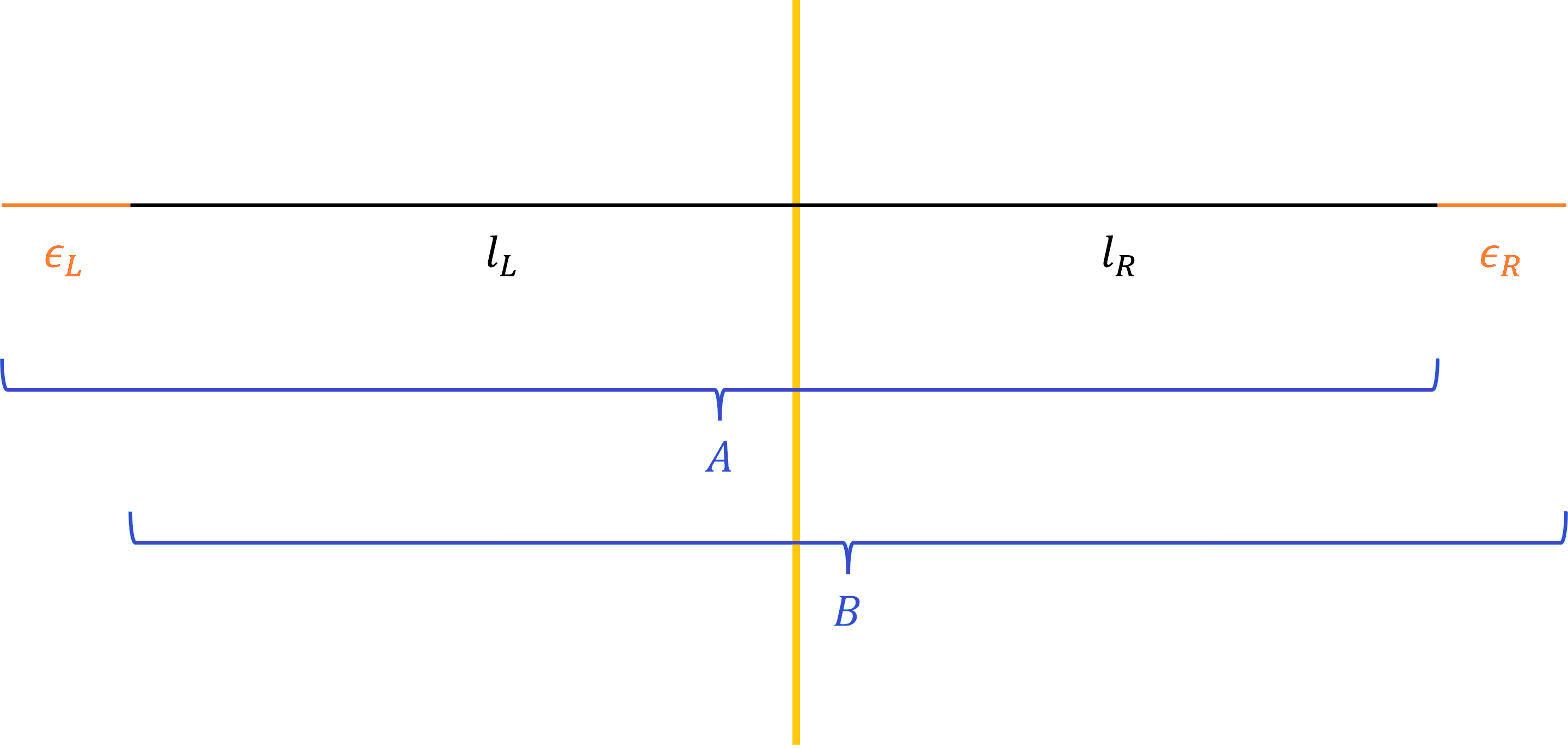}
    \caption{SSA setup used to derive the bounds on $g_{\text{eff}}$.}
    \label{fig:ICFT}
\end{figure}

We can express the EE for an ICFT for arbitrary $l_L$ and $l_R$ as
\begin{equation}
\label{eq::S2}
    S(l_L, l_R) = \frac{c}{6} \log\frac{4 l_L l_R}{\epsilon^2} + F\left( \frac{l_L}{l_R} \right) \ , \quad\quad F\left( \frac{l_L}{l_R}\right)\equiv\log g_{\text{eff}}^{(2)}\left( \frac{l_L}{l_R}\right) \ .
\end{equation}
In order to employ the SSA, we consider the setup in figure \ref{fig:ICFT}
where the length $\epsilon_L, \epsilon_R > 0$. In this case, the SSA yields
\begin{equation}
    S(l_L + \epsilon_L, l_R) + S(l_L, l_R + \epsilon_R) \geq S(l_L, l_R) + S(l_L + \epsilon_L, l_R + \epsilon_R) \ .
\end{equation}
In the limit where $\epsilon_L, \epsilon_R$ become infinitesimally small, we obtain the leading order condition
\begin{equation}
\label{eq::SSAresult}
    \frac{\partial}{\partial l_L} \frac{\partial}{\partial l_R} S(l_L, l_R) \leq 0 \, .
\end{equation}
Substituting (\ref{eq::S1}) into (\ref{eq::SSAresult}) gives
\begin{equation}
    -\frac{1}{l_R^2} F' \left( \frac{l_L}{l_R} \right) - \frac{l_L}{l_R^3} F'' \left( \frac{l_L}{l_R} \right) \leq 0 \, ,
\end{equation}
or equivalently
\begin{equation}
    F'(\rho) + \rho F''(\rho) \geq 0 \, ,
\end{equation}
where we have introduced $\rho=l_L/l_R$. 
On integrating this, it follows that for $0 < \rho \leq 1$,
\begin{equation}
\label{eq::geff2inequality}
   \lim_{\rho \to 0} \left( \rho F'(\rho) \right) \leq \rho F'(\rho) \leq F'(1) \, .
\end{equation}
Defining $c_0 = \lim_{\rho \to 0} \left( \rho F'(\rho) \right)$, the former inequality is
\begin{equation}
\label{eq::geff2bound1}
    \frac{d \log g_{\text{eff}}^{(2)}}{d \rho} \geq \frac{c_0}{\rho} \, ,
    \qquad\qquad
    \text{for } 0 < \rho \leq 1 \, .  
\end{equation}
Now, we further analyze the latter inequality.
The entanglement entropy is symmetric under the exchange of the left and right, so $F(\rho) = F(1/\rho)$ and
\begin{equation}
    F'(\rho) = - \frac{1}{\rho^2} F'(1/\rho) \, .
\end{equation}
Thus, $F'(1)=0$ and we get the $g_{\text{eff}}$-theorem
\begin{equation}
\label{eq::geff2thm}
    \frac{d \log g_{\text{eff}}^{(2)}}{d \rho} \leq 0  \, ,
    \qquad\qquad
    \text{for } 0 < \rho \leq 1 \, .
\end{equation}
These bounds in (\ref{eq::geff2bound1}) and (\ref{eq::geff2thm}) are trivially saturated by the decoupled BCFT answer
\begin{equation}
    F(\rho) = 0 \, ,
\end{equation}
for which $c_0 = 0$.
Thus, these bounds, particularly the $g_{\text{eff}}$-theorem, are tight.
Lastly, we can integrate (\ref{eq::geff2thm}) to get
\begin{equation}
    g_{\text{eff}}^{(2)}(\rho) \geq g \, ,
    \qquad\qquad
    \text{for } 0 < \rho \leq 1 \, ,
\end{equation}
as $g_{\text{eff}}^{(2)}(\rho = 1) = g$.

\subsection{Entropic bounds on \texorpdfstring{$g_{\text{eff}}^{(1)}$}{g-effective-1}} \label{sec::geff1bounds}

We can obtain equivalent bounds on $g_{\text{eff}}^{(1)}$ by starting from the EE expression
\begin{equation}
\label{eq::S1}
    S(l_L, l_R) = \frac{c}{3} \log\frac{l_L + l_R}{\epsilon} + \widetilde{F}\left( \frac{l_L}{l_R} \right) \ , \quad\quad \widetilde{F}\left( \frac{l_L}{l_R}\right)\equiv\log g_{\text{eff}}^{(1)}\left( \frac{l_L}{l_R}\right) \ .
\end{equation}
As $g_{\text{eff}}^{(1)}$ and $g_{\text{eff}}^{(2)}$ are related via (\ref{eq:geff1geff2conv}), we have
\begin{equation}
    \widetilde{F}(\rho) = F(\rho) + \frac{c}{6} \log \frac{4 \rho}{(1+\rho)^2} \, .
\end{equation}
Using (\ref{eq::geff2inequality}), we immediately have the bounds for $0 < \rho \leq 1$,
\begin{equation}
\label{eq::geff1bounds}
    \frac{\widetilde{c}_0}{\rho} - \frac{c}{3} \frac{1}{1 + \rho} \leq \widetilde{F}'(\rho) \leq \frac{c}{6} \frac{(1 - \rho)}{\rho (1 + \rho)} \, ,
\end{equation}
where
\begin{equation}
    \widetilde{c}_0 = \lim_{\rho \to 0} \left( \rho \widetilde{F}'(\rho) \right) = c_0 + \frac{c}{6} \, .
\end{equation}

\subsection{Holographic results}

In this section, we verify the $g_{\text{eff}}$-theorem within the holographic framework. To demonstrate this, we utilize the general expression for the interface entropy (\ref{eq::interfaceentropy}) in conjunction with the relation between the integration constant $c_s$ and the interval ratio $l_L/l_R$ (\ref{eq::relationcsandratio}).
Using (\ref{eq::interfaceentropy}), the derivative of $\log{g_{\text{eff}}^{(2)}}$ with respect to $c_s$ is
\begin{multline}
    \frac{d \log{g_{\text{eff}}^{(2)}}}{d c_s}
    = \lim_{\epsilon\rightarrow0}\Bigg[ \frac{1}{4 G_N} \Bigg( \int_{-r_c^-}^{r_c^+}\frac{c_s e^{A(r)}}{(e^{2A(r)}-c_s^2)^{3/2}}dr + \frac{e^{A(r_c^+)}}{\sqrt{e^{2A(r_c^+)}-c_s^2}} \frac{d r_c^+}{d c_s} \\
    + \frac{e^{A(-r_c^-)}}{\sqrt{e^{2A(-r_c^-)}-c_s^2}} \frac{d r_c^-}{d c_s} - \frac{d \log (2l_L/\epsilon)}{d c_s} -\frac{d \log (2l_R/\epsilon)}{d c_s} \Bigg) \Bigg] \ .
\end{multline}
Using $e^{A(r)} = \cosh r$ as $r \to \pm \infty$ along with the values of $r_c^{\pm}$ in (\ref{eq::rcpm_logs}), we see that second and third terms cancels the fourth and fifth terms respectively. Thus,  
\begin{equation}
    \frac{d \log{g_{\text{eff}}^{(2)}}}{d c_s} = \frac{1}{4 G_N} \int_{-\infty}^{\infty}\frac{c_s e^{A(r)}}{(e^{2A(r)}-c_s^2)^{3/2}}dr \geq 0 \ .
\end{equation}
It is worth noting that if $A(r)$ is symmetric about a certain value of $r$ then $l_L \times l_R$ remains unchanged as $c_s$ is varied, for instance see \cite{Karch:2021qhd} for the Janus case and (\ref{eq::superJanuslR})-(\ref{eq::superJanuslL}) for the super-Janus case.
This further justifies the naturalness of studying $g_{\text{eff}}^{(2)}$ in the holographic setting. 
Taking a derivative of (\ref{eq::relationcsandratio}) with respect to $c_s$, we obtain
\begin{equation}\label{eq::derivativerelationcsratio}
    \frac{d\rho}{d c_s} = \rho \times \left( - \int_{-\infty}^{\infty} \frac{e^{A(r)}}{\left( e^{2A(r)} -c_s^2\right)^{3/2}} dr \right) < 0 \ .
\end{equation}
It follows that 
\begin{equation}
     \frac{d \log{g_{\text{eff}}^{(2)}}}{d\rho} = - \frac{c}{6} \frac{c_s(\rho)}{\rho} \leq 0 \ ,
\end{equation}
which verifies the $g_{\text{eff}}$-theorem (\ref{eq::geff2thm}) in a holographic setup.

\subsection{Free fermion results}

\begin{figure}
    \centering
    \includegraphics[width=0.9\linewidth]{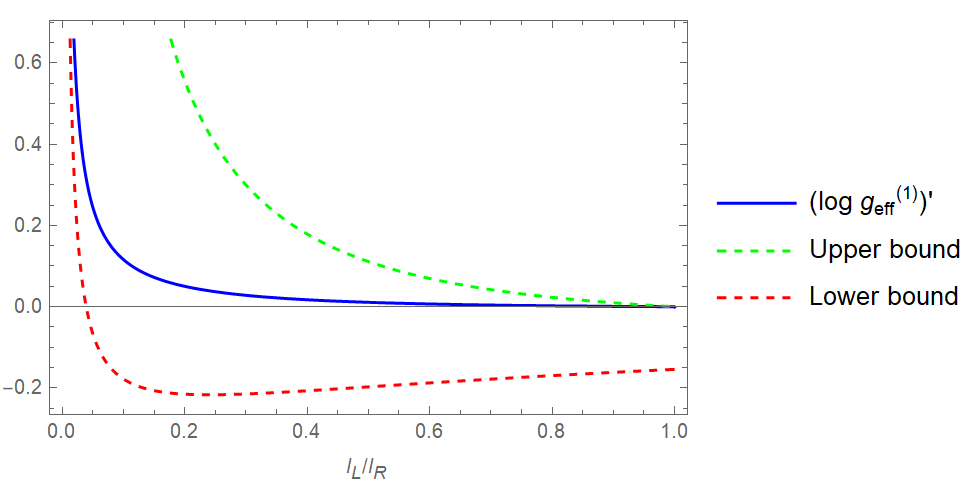}
    \caption{Verification of the constraints on $g_{\text{eff}}^{(1)}$ for a free fermion ICFT with reflection coefficient $r = 0.1$.}
    \label{fig::geff1fermion}
\end{figure}

In this section, we verify the bounds on $g_{\text{eff}}$ in the free fermion theory with a semitransparent interface using the results in \cite{Kruthoff:2021vgv}.
First, we consider an interface that is almost fully transmitting with a small reflection coefficient $r$.
In this case, we have
\begin{equation}
    \log g_{\text{eff}}^{(1)}(\rho) = \frac{r}{8} \left( 1 + \frac{1+\rho^2}{1-\rho^2} \log\rho \right) + O(r^2) \, .
\end{equation}
It is easy to verify that $\log g_{\text{eff}}^{(1)}(\rho) \leq 0$ which represents the aforementioned reduction in entanglement coming from the interface more reflective compared to an ordinary CFT. One can also check that this result satisfies the bounds in (\ref{eq::geff1bounds}), for instance see figure \ref{fig::geff1fermion}.

\begin{figure}
    \centering
    \includegraphics[width=0.9\linewidth]{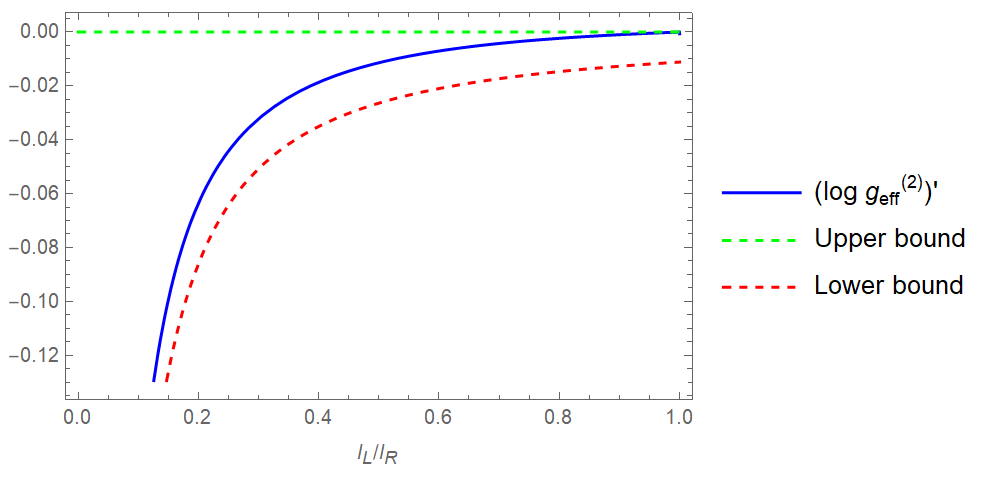}
    \caption{Verification of the constraints on $g_{\text{eff}}^{(2)}$ for a free fermion ICFT with transmission coefficient $t = 0.1$.}
    \label{fig::geff2fermion}
\end{figure}

Next, we consider an interface that is almost fully reflecting with a small transmission coefficient $t$.
In this case, we have
\begin{equation}
    \log g_{\text{eff}}^{(2)}(\rho) = \frac{t}{8} \left( 1 + \frac{1-6\rho+\rho^2}{2\sqrt{\rho}(1-\rho)} \arctan \frac{1-\rho}{2 \sqrt{\rho}} \right) + O(t^2) \, .
\end{equation}
It is easy to verify that $\log g_{\text{eff}}^{(2)}(\rho) \geq 0$ which represents the aforementioned increase in entanglement coming from the interface being more transmissive than two decoupled BCFTs. One can also check that this result satisfies the $g_{\text{eff}}$-theorem in (\ref{eq::geff2thm}), for instance see figure \ref{fig::geff2fermion}.
In this case, the lower bound in (\ref{eq::geff2bound1})  is invalid as the limit computing $c_0$ is ill-defined. Nevertheless, we can modify the derivation to obtain the lower bound
\begin{equation}\label{eq::lowerboundgeff2}
    \frac{d \log g_{\text{eff}}^{(2)}}{d \rho} \geq - \frac{t}{8} \left( \frac{\pi}{8 \rho^{3/2}} + \frac{1}{2 \rho} \right) + O(t^2) \, ,
    \qquad\qquad
    \text{for } 0 < \rho \leq 1 \, .
\end{equation}
This lower bound is also shown in figure \ref{fig::geff2fermion}. Notice that the bound in (\ref{eq::lowerboundgeff2}) is different than the one in (\ref{eq::geff2bound1}), since the former is $\rho$ dependent.

\acknowledgments

We thank Hare Krishna and Ignacio Carreño Bolla for useful discussions. The work of EA is supported by the Severo Ochoa fellowship PA-23-BP22-170 and partially supported by the AEI and the MCIU through the Spanish grant PID2021-123021NB-I00. The work of AK is supported in part by the U.S. Department of Energy under Grant No. DE-SC0022021 and a grant from the Simons Foundation (Grant 651678, AK).
The work of CM is supported in part by the U.S. Department of Energy under Grant No. DE-SC0013528 and the QuantISED Grant No. DE-SC0020360.
EA would like to thank the Weinberg Institute for the kind hospitality.

\appendix
\section{Entropic quantities under KK compactification}
\label{Appendix::KK}

In this appendix, we verify the statement that the 6D computations for the effective central charge $c_{\text{eff}}$ \cite{Gutperle:2015hcv} and the $g-$number \cite{Chiodaroli:2010ur} of super-Janus agree with the 3D computations in the KK reduced theory, first introduced in \cite{Baig:2024hfc}, which we have used. 

\subsection{Boundary entropy}
The 6D $g-$number has been computed in \cite{Chiodaroli:2010ur} and is given by
\begin{equation}\label{6D_g-number}
    S^{6D}_{\text{bdy}}=\frac{16\pi^2 L^2}{4G_6}\cosh^2{\psi}\cosh^2{\theta}\log{(\cosh{\psi}\cosh{\theta})}  \ ,
\end{equation}
where $G_6$ is the six-dimensional Newton constant. We aim to rederive this result from a three-dimensional perspective. The KK compactification of the ten-dimensional super-Janus down to three dimensions was carried out in \cite{Baig:2024hfc}, where the three-dimensional metric is given by
\begin{equation}
    ds^2_{(3)}=R^2\left(\frac{\cosh^2{(r+\psi)}}{\cosh^2{\theta}\cosh^2{\psi}}ds^2_{AdS_2}+dr^2 \right) \ ,
\end{equation}
with $R^2=2L\cosh{\psi}\cosh{\theta}$ is the $AdS_3$ radius. One needs to be careful with the regulated limits on the boundary. We obtain 
\begin{equation}
    \lim_{r\rightarrow \pm \infty}\frac{\cosh^2{(r+\psi)}}{\cosh^2{\theta}\cosh^2{\psi}}=\frac{e^{\pm2(r+\psi)}}{4\cosh^2{\psi}\cosh^2{\theta}}\implies \lambda_{\pm}=\frac{e^{\pm2\psi}}{4\cosh^2{\psi}\cosh^2{\theta}} \ .
\end{equation}
We take $z=z_0$ and the length of the geodesic is given by
\begin{equation}
    \Gamma(\theta,\psi)=\int ds=R\int dx=R(r_{\infty}-r_{-\infty}) \ ,
\end{equation}
with the regulated integration limits given by
\begin{align}
    r_{\infty}&=-\log{\epsilon}-\frac{1}{2}\log{\lambda_+}+\log{2z_0} \\
    r_{-\infty}&=\log{\epsilon}+\frac{1}{2}\log{\lambda_-}-\log{2z_0} \ .
\end{align}
Then
\begin{equation}
    \frac{\Gamma(\theta,\psi)}{R}=-2\log{\epsilon}+2\log(\cosh{\psi}\cosh{\theta})+2\log{2z_0} \ .
\end{equation}
In order to compute the boundary entropy $g$, one needs to subtract the pure AdS contribution from the boundary entropy containing the interface. Therefore we have 
\begin{align}
    S^{3D}_{bdy}&=\frac{\Gamma(\theta,\psi)-\Gamma(0,0)}{4 G_3}=\frac{2R}{4 G_3}\log{(\cosh{\psi}\cosh{\theta})} \nonumber \\
    &=\frac{4 \pi^2 R^4}{4 G_6}\log{(\cosh{\psi}\cosh{\theta})} \ ,
\end{align}
which is in agreement with (\ref{6D_g-number}). Here we have used the fact that the six-dimensional Newton constant is related to the three-dimensional one by $G_6=G_3 V_3=G_3 \times 2\pi^2R^3$, with $V_3$ being the volume of the compactified three-dimensional space.

\subsection{Effective central charge}

The six-dimensional effective central charge $c_{\text{eff}}$ has been computed in \cite{Gutperle:2015hcv} and is given by
\begin{equation}\label{6D_ceff}
    S^{6D}=\frac{1}{\cosh{\psi}\cosh{\theta}}\frac{3\times 32 \pi^3 V_{M_4}L^2\cosh^2{\theta}\cosh^2{\psi}}{\kappa^2_{10}}\frac{1}{6}\log{\left(\frac{l}{\epsilon}\right)} \ ,
\end{equation}
with Newton constant expressed as $\kappa^2_D=8 \pi G^N_D$. Using the KK compactified three-dimensional metric as in \cite{Baig:2024hfc}, we obtain for the length of the geodesic
\begin{equation}
    \Gamma(\theta,\psi)=\int ds =R\int dz\sqrt{\frac{\cosh^2{(r+\psi)}}{\cosh^2{\theta}\cosh^2{\psi}}\frac{1}{z^2}+(r')^2} \ .
\end{equation}
The equations of motion admit a simple solution $r(z)=r_0$. The induced metric on this solution is given by
\begin{equation}
    ds^2_{ind}=R^2\frac{\cosh^2{(r_0+\psi)}}{\cosh^2{\theta}\cosh^2{\psi}}\frac{dz^2}{z^2} \ .
\end{equation}
Therefore, the length of the geodesic
\begin{equation}
    \Gamma(\theta,\psi)=\frac{R\cosh{(r_0+\psi)}}{\cosh{\theta}\cosh{\psi}}\log{\left( \frac{l}{\epsilon}\right)} \ ,
\end{equation}
which is minimized at $r_0=-\psi$. Therefore, the holographic entropy can be computed 
\begin{align}
    S^{3D}&=\frac{\Gamma(\theta,\psi)}{4 G_3}=\frac{R}{4\cosh{\theta}\cosh{\psi} G_3}\log{\left( \frac{l}{\epsilon}\right)} \nonumber \\
    &=\frac{R^42\pi^2 V_{M_4}}{4\cosh{\theta}\cosh{\psi} G_{10}}\log{\left( \frac{l}{\epsilon}\right)} \nonumber \\
    &=\frac{1}{\cosh{\theta}\cosh{\psi}}\frac{L^2 \cosh^2{\psi}\cosh^2{\theta}2\pi^2V_{M_4}8\pi}{\kappa^2_{10}}\log{\left( \frac{l}{\epsilon}\right)} \nonumber \\
    &=\frac{1}{\cosh{\theta}\cosh{\psi}}\frac{3\times 32 \pi^3V_{M_4}\cosh^2{\theta}\cosh^2{\psi}}{\kappa^2_{10}}\frac{1}{6}\log{\left( \frac{l}{\epsilon}\right)} \ ,
\end{align}
which agrees with the six-dimensional expression (\ref{6D_ceff}). Here we have employed the relation of the ten-dimensional Newton constant to the three-dimensional one, given by  $G_{10}=G_3 V_{M_4} \times 2\pi^2 R^3$, with $V_{M_4}$ being the volume of the four-dimensional internal manifold.

 \bibliographystyle{JHEP}
 \bibliography{biblio3.bib}

\providecommand{\href}[2]{#2}\begingroup\raggedright\begin{thebibliography}{10}

\bibitem{Cardy:1986gw}
J.L.~Cardy, \emph{{Effect of Boundary Conditions on the Operator Content of Two-Dimensional Conformally Invariant Theories}}, \href{https://doi.org/10.1016/0550-3213(86)90596-1}{\emph{Nucl. Phys. B} {\bfseries 275} (1986) 200}.

\bibitem{Bachas:2001vj}
C.~Bachas, J.~de~Boer, R.~Dijkgraaf and H.~Ooguri, \emph{{Permeable conformal walls and holography}}, \href{https://doi.org/10.1088/1126-6708/2002/06/027}{\emph{JHEP} {\bfseries 06} (2002) 027} [\href{https://arxiv.org/abs/hep-th/0111210}{{\ttfamily hep-th/0111210}}].

\bibitem{Quella:2006de}
T.~Quella, I.~Runkel and G.M.T.~Watts, \emph{{Reflection and transmission for conformal defects}}, \href{https://doi.org/10.1088/1126-6708/2007/04/095}{\emph{JHEP} {\bfseries 04} (2007) 095} [\href{https://arxiv.org/abs/hep-th/0611296}{{\ttfamily hep-th/0611296}}].

\bibitem{Meineri:2019ycm}
M.~Meineri, J.~Penedones and A.~Rousset, \emph{{Colliders and conformal interfaces}}, \href{https://doi.org/10.1007/JHEP02(2020)138}{\emph{JHEP} {\bfseries 02} (2020) 138} [\href{https://arxiv.org/abs/1904.10974}{{\ttfamily 1904.10974}}].

\bibitem{Calabrese:2009qy}
P.~Calabrese and J.~Cardy, \emph{{Entanglement entropy and conformal field theory}}, \href{https://doi.org/10.1088/1751-8113/42/50/504005}{\emph{J. Phys. A} {\bfseries 42} (2009) 504005} [\href{https://arxiv.org/abs/0905.4013}{{\ttfamily 0905.4013}}].

\bibitem{Affleck:1991tk}
I.~Affleck and A.W.W.~Ludwig, \emph{{Universal noninteger `ground state degeneracy' in critical quantum systems}}, \href{https://doi.org/10.1103/PhysRevLett.67.161}{\emph{Phys. Rev. Lett.} {\bfseries 67} (1991) 161}.

\bibitem{Ryu:2006ef}
S.~Ryu and T.~Takayanagi, \emph{{Aspects of Holographic Entanglement Entropy}}, \href{https://doi.org/10.1088/1126-6708/2006/08/045}{\emph{JHEP} {\bfseries 08} (2006) 045} [\href{https://arxiv.org/abs/hep-th/0605073}{{\ttfamily hep-th/0605073}}].

\bibitem{Azeyanagi:2007qj}
T.~Azeyanagi, A.~Karch, T.~Takayanagi and E.G.~Thompson, \emph{{Holographic calculation of boundary entropy}}, \href{https://doi.org/10.1088/1126-6708/2008/03/054}{\emph{JHEP} {\bfseries 03} (2008) 054} [\href{https://arxiv.org/abs/0712.1850}{{\ttfamily 0712.1850}}].

\bibitem{Karch:2000gx}
A.~Karch and L.~Randall, \emph{{Open and closed string interpretation of SUSY CFT's on branes with boundaries}}, \href{https://doi.org/10.1088/1126-6708/2001/06/063}{\emph{JHEP} {\bfseries 06} (2001) 063} [\href{https://arxiv.org/abs/hep-th/0105132}{{\ttfamily hep-th/0105132}}].

\bibitem{Sakai:2008tt}
K.~Sakai and Y.~Satoh, \emph{{Entanglement through conformal interfaces}}, \href{https://doi.org/10.1088/1126-6708/2008/12/001}{\emph{JHEP} {\bfseries 12} (2008) 001} [\href{https://arxiv.org/abs/0809.4548}{{\ttfamily 0809.4548}}].

\bibitem{Brehm:2015lja}
E.M.~Brehm and I.~Brunner, \emph{{Entanglement entropy through conformal interfaces in the 2D Ising model}}, \href{https://doi.org/10.1007/JHEP09(2015)080}{\emph{JHEP} {\bfseries 09} (2015) 080} [\href{https://arxiv.org/abs/1505.02647}{{\ttfamily 1505.02647}}].

\bibitem{Karch:2022vot}
A.~Karch and M.~Wang, \emph{{Universal behavior of entanglement entropies in interface CFTs from general holographic spacetimes}}, \href{https://doi.org/10.1007/JHEP06(2023)145}{\emph{JHEP} {\bfseries 06} (2023) 145} [\href{https://arxiv.org/abs/2211.09148}{{\ttfamily 2211.09148}}].

\bibitem{Karch:2023evr}
A.~Karch, Y.~Kusuki, H.~Ooguri, H.-Y.~Sun and M.~Wang, \emph{{Universality of effective central charge in interface CFTs}}, \href{https://doi.org/10.1007/JHEP11(2023)126}{\emph{JHEP} {\bfseries 11} (2023) 126} [\href{https://arxiv.org/abs/2308.05436}{{\ttfamily 2308.05436}}].

\bibitem{Karch:2024udk}
A.~Karch, Y.~Kusuki, H.~Ooguri, H.-Y.~Sun and M.~Wang, \emph{{Universal Bound on Effective Central Charge and Its Saturation}}, \href{https://doi.org/10.1103/PhysRevLett.133.091604}{\emph{Phys. Rev. Lett.} {\bfseries 133} (2024) 091604} [\href{https://arxiv.org/abs/2404.01515}{{\ttfamily 2404.01515}}].

\bibitem{Karch:2021qhd}
A.~Karch, Z.-X.~Luo and H.-Y.~Sun, \emph{{Universal relations for holographic interfaces}}, \href{https://doi.org/10.1007/JHEP09(2021)172}{\emph{JHEP} {\bfseries 09} (2021) 172} [\href{https://arxiv.org/abs/2107.02165}{{\ttfamily 2107.02165}}].

\bibitem{Chapman:2018bqj}
S.~Chapman, D.~Ge and G.~Policastro, \emph{{Holographic Complexity for Defects Distinguishes Action from Volume}}, \href{https://doi.org/10.1007/JHEP05(2019)049}{\emph{JHEP} {\bfseries 05} (2019) 049} [\href{https://arxiv.org/abs/1811.12549}{{\ttfamily 1811.12549}}].

\bibitem{Anous:2022wqh}
T.~Anous, M.~Meineri, P.~Pelliconi and J.~Sonner, \emph{{Sailing past the End of the World and discovering the Island}}, \href{https://doi.org/10.21468/SciPostPhys.13.3.075}{\emph{SciPost Phys.} {\bfseries 13} (2022) 075} [\href{https://arxiv.org/abs/2202.11718}{{\ttfamily 2202.11718}}].

\bibitem{Kruthoff:2021vgv}
J.~Kruthoff, R.~Mahajan and C.~Murdia, \emph{{Free fermion entanglement with a semitransparent interface: the effect of graybody factors on entanglement islands}}, \href{https://doi.org/10.21468/SciPostPhys.11.3.063}{\emph{SciPost Phys.} {\bfseries 11} (2021) 063} [\href{https://arxiv.org/abs/2106.10287}{{\ttfamily 2106.10287}}].

\bibitem{Bak:2007jm}
D.~Bak, M.~Gutperle and S.~Hirano, \emph{{Three dimensional Janus and time-dependent black holes}}, \href{https://doi.org/10.1088/1126-6708/2007/02/068}{\emph{JHEP} {\bfseries 02} (2007) 068} [\href{https://arxiv.org/abs/hep-th/0701108}{{\ttfamily hep-th/0701108}}].

\bibitem{Bak:2003jk}
D.~Bak, M.~Gutperle and S.~Hirano, \emph{{A Dilatonic deformation of AdS(5) and its field theory dual}}, \href{https://doi.org/10.1088/1126-6708/2003/05/072}{\emph{JHEP} {\bfseries 05} (2003) 072} [\href{https://arxiv.org/abs/hep-th/0304129}{{\ttfamily hep-th/0304129}}].

\bibitem{Karch:2000ct}
A.~Karch and L.~Randall, \emph{{Locally localized gravity}}, \href{https://doi.org/10.1088/1126-6708/2001/05/008}{\emph{JHEP} {\bfseries 05} (2001) 008} [\href{https://arxiv.org/abs/hep-th/0011156}{{\ttfamily hep-th/0011156}}].

\bibitem{Baig:2024hfc}
S.A.~Baig, A.~Karch and M.~Wang, \emph{{Transmission coefficient of super-Janus solution}}, \href{https://doi.org/10.1007/JHEP10(2024)235}{\emph{JHEP} {\bfseries 10} (2024) 235} [\href{https://arxiv.org/abs/2408.00059}{{\ttfamily 2408.00059}}].

\bibitem{Ryu:2006bv}
S.~Ryu and T.~Takayanagi, \emph{{Holographic derivation of entanglement entropy from AdS/CFT}}, \href{https://doi.org/10.1103/PhysRevLett.96.181602}{\emph{Phys. Rev. Lett.} {\bfseries 96} (2006) 181602} [\href{https://arxiv.org/abs/hep-th/0603001}{{\ttfamily hep-th/0603001}}].

\bibitem{Rangamani:2016dms}
M.~Rangamani and T.~Takayanagi, \emph{{Holographic Entanglement Entropy}}, vol.~931, Springer (2017), \href{https://doi.org/10.1007/978-3-319-52573-0}{10.1007/978-3-319-52573-0}, [\href{https://arxiv.org/abs/1609.01287}{{\ttfamily 1609.01287}}].

\bibitem{Maldacena:1997re}
J.M.~Maldacena, \emph{{The Large $N$ limit of superconformal field theories and supergravity}}, \href{https://doi.org/10.4310/ATMP.1998.v2.n2.a1}{\emph{Adv. Theor. Math. Phys.} {\bfseries 2} (1998) 231} [\href{https://arxiv.org/abs/hep-th/9711200}{{\ttfamily hep-th/9711200}}].

\bibitem{Witten:1998qj}
E.~Witten, \emph{{Anti de Sitter space and holography}}, \href{https://doi.org/10.4310/ATMP.1998.v2.n2.a2}{\emph{Adv. Theor. Math. Phys.} {\bfseries 2} (1998) 253} [\href{https://arxiv.org/abs/hep-th/9802150}{{\ttfamily hep-th/9802150}}].

\bibitem{Lieb:1973zz}
E.H.~Lieb and M.B.~Ruskai, \emph{{A Fundamental Property of Quantum-Mechanical Entropy}}, \href{https://doi.org/10.1103/PhysRevLett.30.434}{\emph{Phys. Rev. Lett.} {\bfseries 30} (1973) 434}.

\bibitem{Headrick:2007km}
M.~Headrick and T.~Takayanagi, \emph{{A Holographic proof of the strong subadditivity of entanglement entropy}}, \href{https://doi.org/10.1103/PhysRevD.76.106013}{\emph{Phys. Rev. D} {\bfseries 76} (2007) 106013} [\href{https://arxiv.org/abs/0704.3719}{{\ttfamily 0704.3719}}].

\bibitem{Harper:2024aku}
J.~Harper, H.~Kanda, T.~Takayanagi and K.~Tasuki, \emph{{g Theorem from Strong Subadditivity}}, \href{https://doi.org/10.1103/PhysRevLett.133.031501}{\emph{Phys. Rev. Lett.} {\bfseries 133} (2024) 031501} [\href{https://arxiv.org/abs/2403.19934}{{\ttfamily 2403.19934}}].

\bibitem{DHoker:2007zhm}
E.~D'Hoker, J.~Estes and M.~Gutperle, \emph{{Exact half-BPS Type IIB interface solutions. I. Local solution and supersymmetric Janus}}, \href{https://doi.org/10.1088/1126-6708/2007/06/021}{\emph{JHEP} {\bfseries 06} (2007) 021} [\href{https://arxiv.org/abs/0705.0022}{{\ttfamily 0705.0022}}].

\bibitem{Chiodaroli:2009yw}
M.~Chiodaroli, M.~Gutperle and D.~Krym, \emph{{Half-BPS Solutions locally asymptotic to AdS(3) X S**3 and interface conformal field theories}}, \href{https://doi.org/10.1007/JHEP02(2010)066}{\emph{JHEP} {\bfseries 02} (2010) 066} [\href{https://arxiv.org/abs/0910.0466}{{\ttfamily 0910.0466}}].

\bibitem{Chiodaroli:2010ur}
M.~Chiodaroli, M.~Gutperle and L.-Y.~Hung, \emph{{Boundary entropy of supersymmetric Janus solutions}}, \href{https://doi.org/10.1007/JHEP09(2010)082}{\emph{JHEP} {\bfseries 09} (2010) 082} [\href{https://arxiv.org/abs/1005.4433}{{\ttfamily 1005.4433}}].

\bibitem{Takayanagi:2011zk}
T.~Takayanagi, \emph{{Holographic Dual of BCFT}}, \href{https://doi.org/10.1103/PhysRevLett.107.101602}{\emph{Phys. Rev. Lett.} {\bfseries 107} (2011) 101602} [\href{https://arxiv.org/abs/1105.5165}{{\ttfamily 1105.5165}}].

\bibitem{Gutperle:2012hy}
M.~Gutperle and J.~Samani, \emph{{Holographic RG-flows and Boundary CFTs}}, \href{https://doi.org/10.1103/PhysRevD.86.106007}{\emph{Phys. Rev. D} {\bfseries 86} (2012) 106007} [\href{https://arxiv.org/abs/1207.7325}{{\ttfamily 1207.7325}}].

\bibitem{Casini:2004bw}
H.~Casini and M.~Huerta, \emph{{A Finite entanglement entropy and the c-theorem}}, \href{https://doi.org/10.1016/j.physletb.2004.08.072}{\emph{Phys. Lett. B} {\bfseries 600} (2004) 142} [\href{https://arxiv.org/abs/hep-th/0405111}{{\ttfamily hep-th/0405111}}].

\bibitem{Hirata:2006jx}
T.~Hirata and T.~Takayanagi, \emph{{AdS/CFT and strong subadditivity of entanglement entropy}}, \href{https://doi.org/10.1088/1126-6708/2007/02/042}{\emph{JHEP} {\bfseries 02} (2007) 042} [\href{https://arxiv.org/abs/hep-th/0608213}{{\ttfamily hep-th/0608213}}].

\bibitem{Gutperle:2015hcv}
M.~Gutperle and J.D.~Miller, \emph{{Entanglement entropy at holographic interfaces}}, \href{https://doi.org/10.1103/PhysRevD.93.026006}{\emph{Phys. Rev. D} {\bfseries 93} (2016) 026006} [\href{https://arxiv.org/abs/1511.08955}{{\ttfamily 1511.08955}}].

\end{thebibliography}\endgroup

\end{document}